\documentclass[aps,floatfix,onecolumn,a4paper,noshowpacs, nofootinbib,superscriptaddress,11pt]{revtex4}
\usepackage{graphicx,float}\usepackage{graphicx,float}
\usepackage[all]{xy}
\usepackage{amsmath,upgreek,bigints}
\usepackage{amssymb}
\usepackage{color}
\usepackage{epsfig,bm}		
\usepackage{graphicx,epstopdf}
\usepackage{subfigure}
\usepackage{pdfpages}
\usepackage[colorlinks,hyperindex]{hyperref}
 
  \newcommand{\clt}{\textcolor{black}}
    
       \newcommand{\smre}{strange vector kaon  resonances}
        \newcommand{\smrs}{strange vector kaon  radial resonance}
        \newcommand{\smr}{strange vector kaon  resonances\;}
            \newcommand{\sma}{strange vector kaons\;}
              \newcommand{\smas}{strange vector kaon\;}
              \newcommand{\smae}{strange vector kaons}
\definecolor{green1}{RGB}{0,128,0} 
\hypersetup{backref=true,pagebackref=true}
\hypersetup{%
  colorlinks = true,
  linkcolor  = blue,
  citecolor = cyan,
}
\usepackage{bookmark,textgreek}
\usepackage{hyperref,color,xcolor}
\hypersetup{hidelinks,hyperindex=true,colorlinks=true,breaklinks=true,urlcolor= blue}
\hypersetup{%
  colorlinks = true,
  linkcolor  = blue
}\usepackage{amssymb}
\newsavebox{\foobox}

\usepackage{graphicx,float,tikz}
\usepackage[all]{xy}
\newcommand\ringring[1]{%
  {% make an Ord atom
   \mathop{\kern0pt #1}\limits^{% set a box over the variable
     \vbox to-1.85ex{
       \kern-2ex % lower the ring accents
       \hbox to 0pt{\hss\normalfont\kern.1em \r{}\kern-.45em \r{}\hss}%
       \vss % fill
     }% end of \vbox
   }% end of the superscript
  }% end of \mathop
}\newcommand\orcidroldao{{\href{https://orcid.org/0000-0003-3978-532X}{\orcidicon}}}
\newcommand{\orcidicon}{%
	\begin{tikzpicture}
	\draw[lime, fill=lime] (0,0)
		circle [radius=0.16]
		node[white] {{\fontfamily{qag}\selectfont \tiny ID}};
	\draw[white, fill=white] (-0.0625,0.095)
		circle [radius=0.007];
	\end{tikzpicture}	\hspace{-2mm}
}
\newcommand{\bpartial}{\mathop{\partial\kern -4pt\raisebox{.8pt}{$|$}}}

\newcommand{\bes}{\begin{subequations}}
\newcommand{\ees}{\end{subequations}}
\def\beq{\begin{eqnarray}}

\def\eeq{\end{eqnarray}}
\def\be{\begin{equation}}
\def\ee{\end{equation}}

\usepackage{natbib}
\setcitestyle{square,numbers}

\begin{document}

\title{Configurational entropy and shape complexity of strange vector kaons in AdS/QCD}
\author{R. da Rocha\orcidroldao\!\!}
\email{roldao.rocha@ufabc.edu.br}
\affiliation{Federal University of ABC, Center of Mathematics, Santo Andr\'e, 09580-210, Brazil}
\author{P. H. O. da Silva}
\email{silva.pedro@ufabc.edu.br}
\affiliation{Federal University of ABC, Center of Physics, Santo Andr\'e, 09580-210, Brazil}

\begin{abstract}
The mass spectrum of vector kaons radial resonances is scrutinized, using  AdS/QCD with a deformed dilaton  that arises from the constituent quark masses. Both the differential configurational entropy and the differential configurational complexity are computed and used to achieve the mass spectrum of \smr with higher  radial excitation quantum number. This approach amalgamates AdS/QCD and the experimental mass spectrum of already detected strange vector kaons in the Particle Data Group, providing a hybrid technique to study the next generation of \smre.

 \end{abstract}
\maketitle

\section{Introduction}

Shannon's information entropy of a  measurable probability function  evaluates the possible results that come out of a  random variable. This concept is fundamental for defining configurational information measures (CIMs). One particular CIM is the differential configurational entropy (DCE), which evaluates the information entropy of a continuous  system and regulates the way how the system constituents are correlated \cite{Gleiser:2018kbq,Gleiser:2018jpd}.  The DCE  estimates how information, representing a physical system, can be allocated into wave modes, in the momentum space. The DCE also renders to the very limit of lossless compression of data, for analogical sources of information. 
Ref. \cite{Bernardini:2016hvx}  was the first approach to employing the DCE to scrutinize mesons in AdS/QCD, using the asymptotically AdS metric as the near horizon limit of a  stack of parallel $D_3$ branes, and also applying the Witten--Sakai--Sugimoto brane model as well. This approach was extended to encompass tachyonic fields in AdS/QCD, using also the DCE to acquire an improved mass spectrum of mesons \cite{Barbosa-Cendejas:2018mng}. The DCE has been providing relevant aspects of hadronic matter and important new properties of hadronic resonances. 
Together with the holographic QCD, it yields an accurate quantitative prediction of confined hadronic properties and their mass spectrum  \cite{Bernardini:2018uuy,Braga:2017fsb,Karapetyan:2018oye,Karapetyan:2018yhm,Braga:2018fyc,Braga:2020myi,daRocha:2021imz}, besides also endorsing  well-established hadronic features. The latest applications of the DCE  to AdS/QCD shed new phenomenological light on dozens of light- and heavy-flavor meson families,  
 enclosing quarkonia, exotic mesons, higher-spin tensor mesons, baryonic matter, nuclides, and glueball fields, also including their finite-temperature description, the quark-gluon plasma with a magnetic field, the color-glass condensate, and other various aspects of QCD phenomenology      \cite{Colangelo:2018mrt,Ferreira:2020iry,Braga:2020hhs,Karapetyan:2019fst,Karapetyan:2021vyh,Karapetyan:2020yhs,Ma:2018wtw,Frederico:2014bva,dePaula:2009za,Braga:2021fey,Braga:2021zyi,Karapetyan:2016fai,Karapetyan:2021crv,MartinContreras:2022lxl,Karapetyan:2017edu}. In particular, the results concerning the use of the DCE to study the mass spectroscopy of several meson families have been  already validated by the Particle Data Group (PDG) \cite{pdg}.  Besides, the new results using the DCE also comply with new candidates for meson resonances \cite{pdg}. The DCE has also been employed to study the gravitational collapse and turbulence in AdS \cite{Barreto:2022ohl} and to probe several other several aspects of  AdS/CFT   \cite{Braga:2019jqg,Barreto:2022ohl,Cruz:2019kwh,Lee:2019tod,Bazeia:2018uyg,Bazeia:2021stz,Alves:2017ljt,Braga:2020opg,Fernandes-Silva:2019fez,Lee:2018zmp}. 

Several families of mesonic resonances have been explored in AdS/QCD,
having their mass spectrum obtained in this scenario, as well as other properties, such as the decay width. Although QCD presents  chiral symmetry breaking, and for this reason QCD is not exactly a conformal field theory (CFT), the regime of approximations of AdS/CFT can be still useful, where an AdS  bulk comprises a gravitational sector that is  weakly coupled. The holographic dual CFT lives in the boundary of the AdS bulk. In the AdS/CFT correspondence, physical fields inhabiting the AdS bulk are dual objects to operator fields in QCD, which are strongly coupled and are trapped in the AdS boundary. 
Holographic QCD represents the non-perturbative confined regime of QCD,  where the mesonic fields can be associated with  gauge fields of a Yang--Mills theory with flavor. 

In AdS/QCD, the additional dimension, off the AdS boundary, is essentially  equivalent to the energy scale of QCD \cite{Csaki,Karch:2006pv,Branz:2010ub}. A possible bottom-up approach of AdS/QCD consists of the soft wall,  prescribing Regge trajectories to come out from confinement. The soft wall AdS/QCD is usually implemented by a dilaton that is a quadratic function of the energy scale  \cite{Colangelo:2008us,MartinContreras:2020cyg}.  Bottom-up AdS/QCD  complies with accurately fitting the QCD phenomenology as the asymptotic boundary field theory of gravity coupled to a gauge theory on the AdS bulk. 
Linear Regge trajectories are a reasonable approximation to study mass spectroscopy of light-flavor mesonic states, as long as the radial quantum number is controlled below a critical number,  beyond which the Regge trajectories cannot be approached by a linear fit.  When hadronic states are constituted by strange ($s$), bottom ($b$), charm ($c$), or even top ($t$) quarks, the linear regime $m_n^2\propto n$ typically falls apart  \cite{MartinContreras:2020cyg,Afonin:2014nya}. 
 In addition, due to the mass gap between the up ($u$) and down ($d$) light quarks, compared to the other heavy quarks, one must take into account massive quarks constituents for modeling mesonic states. The $s$ quark is two orders of magnitude more massive than the light quarks, whereas the $c$ and the $b$ quarks are three orders of magnitude more massive than the light quarks. The $t$ quark has mass five orders of magnitude larger than the light quarks. Since our aim in this work consists of studying the \smre, formed by a combination between the $s$ quark and either the $u$ or $d$ quarks, the massive constituent model must be employed \cite{MartinContreras:2020cyg}. Experimental data yields a well-established model to describe strange vector mesons, whose lighter family members have a similar quark composition.  The first possibility about the quark constituents encompasses the $K^*(892)^+=(u\bar{s})$ and the $K^*(892)^-~=~(\bar{u} s)$
charged strange vector mesons, 
%\footnote{One of them is the antiparticle to the other.},
with mass $(891.66\pm0.26)$ MeV/$c^2$, whereas the neutral strange vector kaon $K^*(892)^0=~(d\bar{s})$ has mass $(895.55\pm0.20)$~MeV/$c^2$, agreeing up to 4~MeV/$c^2$ \cite{pdg,ALICE:2021xyh}. 
At LHC energies between 8 and 13 TeV, the measurement of the neutral and charged strange vector mesons is an arduous task. They can be reconstructed from their hadronic decay into a charged pion and a neutral kaon,  for the $K^{*\pm}(892)$, and into a charged pion and a neutral kaon,  for the $K(892)^{*0}$ \cite{ALICE:2021xyh}.  Nonetheless, it is hard to reconstruct them, since kaons and pions undergo final state interactions with other hadrons and can either scatter anew or even be absorbed.  The other two members in the strange vector meson family, which have been already experimentally detected, are the $K^*(1410)$ and the $K^*(1680)$, which together with the $K^*(892)$, form the resonances corresponding to $n=1,2,3$.  The neutral $K^*(1410)^0$ appears in the decay of a strange $B$ meson, $B^0_s = (s\bar{b})$, into a fundamental charmonium $J/\psi = (\bar{c}c)$ and a $K^*(1410)^0$, whose exit channel consists of a $K^\pm$ kaon  and a $\pi^\mp$ pion \cite{pdg}.
Recent experiments reported the $B^0_s$ decay into a second excited charmonium  $\psi(2S)$ and a $K^*(1410)^0$, but with a exit channel formed by  a $K^+$ charged kaon and a $\pi^-$ pion \cite{Li:2019pzx}. 
On the other hand,  the neutral $K^*(1680)^0$ arises in the decay of the strange $B$ meson, $B^0_s$  into the charmonium $J/\psi$ and a $K^*(1680)^0$, whose exit channel consists of a $K^+$ kaon and a $\pi^-$ pion \cite{pdg,Li:2019pzx}. Up to now, there are no experimental results reported that involve the strange vector kaon $K^*(1680)^0$ decaying into a  charmonium  $\psi(2S)$, up to our knowledge. 
Charged strange vector kaons can appear in experiments 
as the decay of either a  strange $B$ meson, $B^0_s$, or a light-heavy $B$ meson, $B^0 = (d\bar{b})$, into a kaon $K^\pm$ and a strange vector kaon resonance $K^*(1410)^\mp$ \cite{Li:2018psm}. Also, depending on the entrance channel energy, a 
$K^*(1680)^\mp$ might be produced, however, this event is reported to be much rarer.  
In the context of AdS/QCD, Ref.  \cite{MartinContreras:2020cyg}, addresses only the neutral  strange vector kaons, 
whose average mass differs around 0.5\% when compared to charged strange vector kaons. However, this slight difference may reflect a distinct mass spectrum of the strange vector kaon  resonances with higher excitation quantum numbers $n$.  Describing both possibilities regarding charged and neutral strange vector kaons allows more range to probe the \smr with higher values of $n$ \cite{Liu:2017xwi}.

Complementary to the DCE, another CIM is the differential configurational complexity (DCC). As the DCE evaluates  the number of bits needed to assemble any field configuration from the wave modes portraying a physical  system, the DCC computes the   complexity of such a construction, measuring the informational complexity intrinsic to the shape of the configuration assumed by the system. When the wave modes coming up with the shape of a field configuration have nearly equal contributions, the DCC reduces, whereas when the wave modes have a non-uniform weight distribution, the DCC increases,  instead \cite{Gleiser:2018jpd}. Recently the DCC and the DCE were successfully  
used to scrutinize topological defects and the membrane paradigm of AdS/CFT 
 \cite{Barreto:2022mbx}. 
In this work, the DCE and DCC of the \smr will be computed in the soft wall AdS/QCD, with a deformed dilaton that takes into account the massive constituent quarks. Both the DCE and the DCC of the \smr  can be written as a  function of the excitation level $n$. The interpolation curves engender the first kind of DCE- and DCC-Regge-like trajectories. When they are expressed instead as a 
function of the strange vector kaons' experimental mass spectrum, a second kind of DCE- and DCC-Regge-like trajectories correspond to the respective interpolation curves. When these Regge-like trajectories are extrapolated, the mass spectrum of \smr with $n>3$ can be obtained, working out the interpolation formula of  DCE- and DCC-Regge-like trajectories \smae. 

This work is organized as follows: Sec. \ref{sec1}   introduces the main ingredients of the bottom-up soft wall AdS/QCD, with a deformed dilaton that encodes the massive quarks that are confined to form strange vector kaons.  
The mass spectrum of strange vector kaon meson resonances is then obtained as a solution of the Schr\"odinger-like equation that is equivalent, after a Bogoliubov transformation, to the equation of motion of a gauge vector field describing the strange vector meson.  
 Sec. \ref{sec2} describes and discusses the DCE protocol. The DCE for both the charged and neutral strange vector kaon meson families is then evaluated first as a function of the radial quantum number of \smre, whose interpolation curve generates the first kind of DCE-Regge-like trajectory. When the DCE  is computed as a function of the mass spectrum of \smre, the associated interpolation curve yields a second kind of DCE-Regge-like trajectory. Concomitant use of the first and second kinds of DCE-Regge-like trajectories yields the mass spectrum of the next generation of \smre.
  Sec. \ref{sec21} implements a similar technique, this time using the DCC.
  New interpretations and the phenomenological comparison between the DCE and DCC methods are addressed. 
 %The analysis and the interpolation of  the resulting data consist, respectively, of the first and second  types of DCE-Regge-like trajectories. 
 Extrapolating the DCC-Regge-like trajectories also allows obtaining the  mass spectrum of heavier strange vector kaon resonances. Besides  potentially matching up to existing further mesonic states in PDG, the results here obtained also direct reasonable values of masses for the next generation of \smr to eventually guide the next runnings in current experiments. Sec. \ref{iv} encloses discussion, analysis, and relevant perspectives.

\section{AdS/QCD fundamentals}
\label{sec1}

Strange vector kaons can be modeled by the soft wall AdS/QCD \cite{Karch:2006pv}, when a bulk gauge vector field  $\mathsf{V}_M = (\mathsf{V}_\mu,\mathsf{V}_z)\,$ is taken into account, where greek letters assume spacetime indexes, $0$ to $3$, and Latin capital letters denote AdS bulk indexes, $0,1,2,3,5$. 
In holographic QCD, the bulk gauge vector field plays the role of a dual object to the current density 
$\mathsf{J}^\mu(x) = \bar{\mathsf{q}}(x)\upgamma^\mu \mathsf{q}(x)$, which lives in the AdS  boundary, for $ \mathsf{q}(x)$  denoting a quark fermionic field. 
The $\{\upgamma^\mu\}$ is the set of Dirac matrices and the Dirac spinor adjoint is the usual one, given by $\bar{\mathsf{q}}(x)=\mathsf{q}^
\dagger(x)\upgamma^0$. 
The bulk gauge vector field is regulated by the following action,  
\begin{equation}
\mathcal{I} =- \frac{1}{\mathsf{g}_5^2}  \int \sqrt{-g}  {\scalebox{.94}{$\mathcal{L}$}}\,d^4xdz,
\label{vectorfieldaction}
\end{equation}
where 
\beq\label{lalag}
{\scalebox{.94}{$\mathcal{L}$}}=\frac14\, e^{-{\scalebox{.68}{$\upphi$}} (z)}  {F^*}^{MN} F_{MN}\eeq
is the Lagrangian density and $F_{MN}$ is the gauge field strength.  
The metric   
\begin{equation}\label{space1}
 ds^2 =g_{MN}dx^Mdx^N= e^{2\mathsf{A}(z)}(\mho_{\mu\nu}dx^\mu dx^\nu+dz^2)\,,
\end{equation}
endows the AdS bulk, where  ${A}(z) = -\ln(z/R)$ is the  warp factor and  $\mho_{\mu\nu}$ denotes the  metric equipping the boundary. A complementary description of vector mesons can be found in Refs. \cite{Bayona:2010bg,BallonBayona:2009ar}. The gauge vector field interacts to the dilaton ${\scalebox{.98}{$\upphi$}}(z)$ in the bulk.  
 Ref. \cite{MartinContreras:2020cyg} introduced the expression
\beq
{\scalebox{.98}{$\upphi$}}(z) = ({\scalebox{.94}{$\kappa$}} z)^{2-{\scalebox{.64}{$\upalpha$}}},\label{devia}
\eeq containing a suitable deformation to the usual quadratic dilatonic field in the soft wall AdS/QCD. The energy parameter  ${\scalebox{.94}{$\kappa$}}$ corresponds to the mass scale in QCD and ${\scalebox{.94}{$\upalpha$}}$ is responsible for departing from the usual  formulation of the quadratic dilaton, deforming it to encompass the constituent model of massive quarks,   forming strange vector kaon fields. To describe \smr in this context, the strange quark enters into the framework and there is no space for the massless description of light-flavor mesons, as usually done when just the $u$ and $d$ quarks are taken into account. When the limit ${\scalebox{.94}{$\upalpha$}}\to0$ is performed, the massless quark constituent model of light-flavor mesons is obtained, as a byproduct of  the   AdS/QCD soft wall with a quadratic dilaton.

When the action (\ref{vectorfieldaction}) is employed, as well as the gauge choice  $\mathsf{V}_z=0$, the following equation of motion regulates the AdS bulk gauge vector field, 
\begin{equation}
\frac{\partial^2\mathsf{V}_\mu(q,z)}{\partial z^2}-B'(z)\frac{\partial\mathsf{V}_\mu(q,z)}{\partial z}-q^2\mathsf{V}_\mu(q,z)=0,    
\end{equation}
\noindent where $B(z)$ is the difference between the dilaton and the warp factor, namely, $B(z)={\scalebox{.98}{$\upphi$}}(z)-\mathsf{A}(z)$, whereas the prime indicates taking the differentiation with respect to AdS bulk coordinate $z$. The bulk gauge field describing the strange vector kaon  can be decomposed when  the  bulk-to-boundary propagator ${\scalebox{.94}{$\upxi$}}(q,z)$ is regarded, as
 \beq\label{qiqi}
\mathsf{V}_\mu(q,z)=\mathsf{V}_\mu(q)\,{\scalebox{.94}{$\upxi$}}(q,z),\eeq where  $\mathsf{V}_{\mu}(q) =
\lim_{z\to 0} \mathsf{V}_\mu (q,z)$ denotes the gauge vector field on the CFT boundary.   Hence, the equation of motion that governs the bulk gauge vector kaon field can be rewritten as a Schr\"odinger-like equation, as long as a Bogoliubov scaling  ${\scalebox{.94}{$\upxi$}}(q,z)\mapsto e^{{\scalebox{.72}{$\upphi$}}(z)/2}\upxi(q,z)$ is taken into account, leading to \cite{MartinContreras:2020cyg} 
\begin{equation}\label{schr}
\left[-{\scalebox{.94}{$\partial$}}_z^2+U(z)\right]\,\upxi(q,z)=-q^2\upxi(q,z),     
\end{equation}
with the general potential 
\begin{eqnarray}\label{potencial11}
U(z)&=&\frac{1}{4}B^{\prime2}(z)-\frac{1}{2} B''(z),
\end{eqnarray}
\noindent  
Eq. (\ref{schr}) yields the mass spectrum $m_n^2 = -q^2=4\kappa^2n$ of \smrs, as  eigenvalues of the Schr\"odinger-like operator, where the index $n$ is the radial excitation quantum number, labeling each one of  the \smae.

To the CFT boundary term $\mathsf{V}_\mu(q)$ in Eq.  \eqref{qiqi}, one must assume the Dirichlet first-type condition $\lim_{z\to0}{\scalebox{.94}{$\upxi$}}(q,z) = 1$. 
 The term $\mathsf{V}_\mu(q)$ plays the role of a source, which is responsible for  generating the 2-point correlators of the current density operator  $  \mathsf{J}^\mu (x^\rho)$, given by  
\begin{eqnarray}
\!\! \langle \,0\, \vert \, \mathsf{J}_\mu (x^\rho) \mathsf{J}_\nu (x^{\prime\rho}) \,  \vert \,0\, \rangle= \frac{\delta}{\delta \mathsf{V}_{\mu}(x^\rho)} \frac{\delta}{\delta \mathsf{V}_{\nu}(x^{\prime\rho})}
e^{- \mathcal{S}},
 \label{2pointfunction}
\end{eqnarray}
\noindent on the CFT boundary, where $\mathcal{S}$  stands for the on-shell action, arising from the boundary expression
 \begin{equation}\label{lalal}
\!\!\!\!\!\!\!\!\mathcal{S}= - \frac{R}{2{\mathsf{g}_5^2}}  \lim_{z \to 0} \int  \frac{1}{z}{e^{-({\scalebox{.74}{$\kappa$}} z)^{2-{\scalebox{.6}{$\upalpha$}}}} \mho^{\mu\nu}\mathsf{V}_\mu {\scalebox{.94}{$\partial$}}_z \mathsf{V}_\nu }\,d^4x.
\end{equation}  The correlator therefore can be expressed as 
 \begin{eqnarray}
  \!\Pi (q^2)  \!=\!   -\frac{R}{{\mathsf{g}_5^2} q^ 2 } \lim_{z \to 0}\left[ \frac{1}{z} e^{ -({\scalebox{.74}{$\kappa$}} z)^{2-{\scalebox{.6}{$\upalpha$}}}}    {\scalebox{.94}{$\upxi$}}(q,z) {\scalebox{.94}{$\partial$}}_z {\scalebox{.94}{$\upxi$}}(q,z) 
\right]. \label{hol2point}
\end{eqnarray}      
 
Substituting the deformed dilaton \eqref{devia} in the Schr\"odinger potential \eqref{potencial11} implies that   \cite{MartinContreras:2020cyg}
\beq\label{ppp}
\!\!\!\!\!\!\!\!\!U({\scalebox{.94}{$\kappa$}},{\scalebox{.94}{$\upalpha$}},z)&=&\left(1-\frac{{\scalebox{.94}{$\upalpha$}}}{2}\right)\frac{{\scalebox{.94}{$\kappa$}} }{z}({\scalebox{.94}{$\kappa$}}  z)^{1-{\scalebox{.6}{$\upalpha$}} }+\left({\scalebox{.94}{$\upalpha$}}-2\right)^2\frac{{\scalebox{.94}{$\kappa$}}^2}{4}({\scalebox{.94}{$\kappa$}} z)^{2-2 {\scalebox{.6}{$\upalpha$}} }+\frac{3}{4 z^2}+\left(\frac{3{\scalebox{.94}{$\upalpha$}}}2-\frac{{\scalebox{.94}{$\upalpha$}} ^2}{2}   -{\scalebox{.94}{$\kappa$}} ^2\right){\scalebox{.94}{$\kappa$}}^2 ({\scalebox{.94}{$\kappa$}}  z)^{-{\scalebox{.6}{$\upalpha$}} }.
\eeq
 The deformation parameter  ${\scalebox{.94}{$\upalpha$}}$ carries the information that within \smr there are now massive quark constituents as well as their flavor  \cite{MartinContreras:2020cyg}. 
To get the radial strange vector kaon resonances, the deformed  potential in Eq. \eqref{ppp} must be substituted into the Schr\"odinger-like equation  (\ref{schr}), whose numerical solutions, for the vector kaon,  are encoded in Tables  \ref{scalarmasses1} and \ref{scalarmasses10}, with slightly different values of the  ${\scalebox{.94}{$\kappa$}}$ and ${\scalebox{.94}{$\upalpha$}}$ parameters, chosen to fit the mass spectrum of the neutral and charged strange vector kaons. The pioneering work in Ref. \cite{MartinContreras:2020cyg} analyzed neutral strange vector mesons and here the discussion about the charge strange vector mesons will be included. Their  quark composition is given by  $K^*(892)^+=(u\bar{s})$, $K^*(892)^0=~(d\bar{s})$, and  
$K^*(892)^-~=~(\bar{u} s)$.
Their experimental masses differ by about 4~MeV/$c^2$, being 
$m_{K^{*\pm}}~=(891.66~\pm~0.26)$ MeV/$c^2$  and $m_{K^{*0}}~=(895.55~\pm~0.20)$~MeV/$c^2$~\cite{pdg}.

In general, the hadronic mass spectra, and the Regge trajectories, are constructed from the eigenvalues of the Schr\"odinger-like equation, which is fixed by the structure of the $B(z)$ function. In the context of the original soft wall model \cite{Karch:2006pv}, the potential is fixed by $B(z) = \kappa^2 z^2 - \ln(R/z)$ obtaining the linear spectrum $m_n^2=4\kappa^2n$
associated with vector mesons with massless constituent quarks, for $\kappa=0.388$ GeV \cite{Karch:2006pv}. If one wants to regard the case when the constituent quarks are massive, Ref. \cite{MartinContreras:2020cyg} studied the behavior of the deformation parameter $\upalpha$ in the dilaton and the energy scale $\kappa$, showing that they vary according to the average  constituent quark mass as 
\beq
\label{alpha-fit}
{\scalebox{.94}{$\upalpha$}}(\bar{m}) &=& 0.8454-0.8485 e^{-0.4233 \bar{m}^2},\\
\label{kappa-fit}
\kappa (\bar{m}) &=& 15.2085-14.8082e^{-0.0524 \bar{m} ^{2}}.
\eeq

 Here, the idea is to include the constituent mass fraction as part of the model to derive the proper pair of parameters ${\scalebox{.94}{$\upalpha$}}$ and $\kappa$ for the charged and neutral strange vector kaon families,  within the constituent massive quark model. 
	In this work, the same set of values for the constituent quark masses will be used, as in Ref. \cite{MartinContreras:2020cyg}. One adopts $m_u=0.336$ GeV/$c^2$, $m_d=0.340$ GeV/$c^2$ and $m_s=0.486$ GeV/$c^2$.

	In Eqs. (\ref{alpha-fit}, \ref{kappa-fit}), the mass $\bar{m}$ is the average of the masses of the pair of quarks that compose the meson. Therefore, one can see that this implies two possibilities for describing strange vector kaons. The first consists  of $d\bar{s}$ or $\bar{d}s$, which respectively models the $K^*(892)^0$ and $\bar{K}^*(892)^0$, which yields  $\bar{m}=413$ MeV/$c^2$ and, consequently,  ${\scalebox{.94}{$\upalpha$}}=0.056$ and $\kappa=532.06$ MeV/$c^2$. 		The second possibility consists of $u\bar{s}$ or $\bar{u}s$, which respectively models the $K^*(892)^+$ and ${K}^*(892)^-$ charged strange vector kaons. For this case, the average mass $\bar{m}=411$ MeV/$c^2$  implies the values ${\scalebox{.94}{$\upalpha$}}=0.0554$ and $\kappa=530.79$ MeV/$c^2$.

Also, it is worth emphasizing another usefulness of the DCE-based AdS/QCD hybrid method presented here. It consists of a feasible alternative to standard bottom-up AdS/QCD predictions, which according to the tables that follow, already provides a relative error per state ({RE}). 
In what follows the respective relative errors per state are displayed.

\begin{figure}[H]
	\centering
	\includegraphics[width=8.9cm]{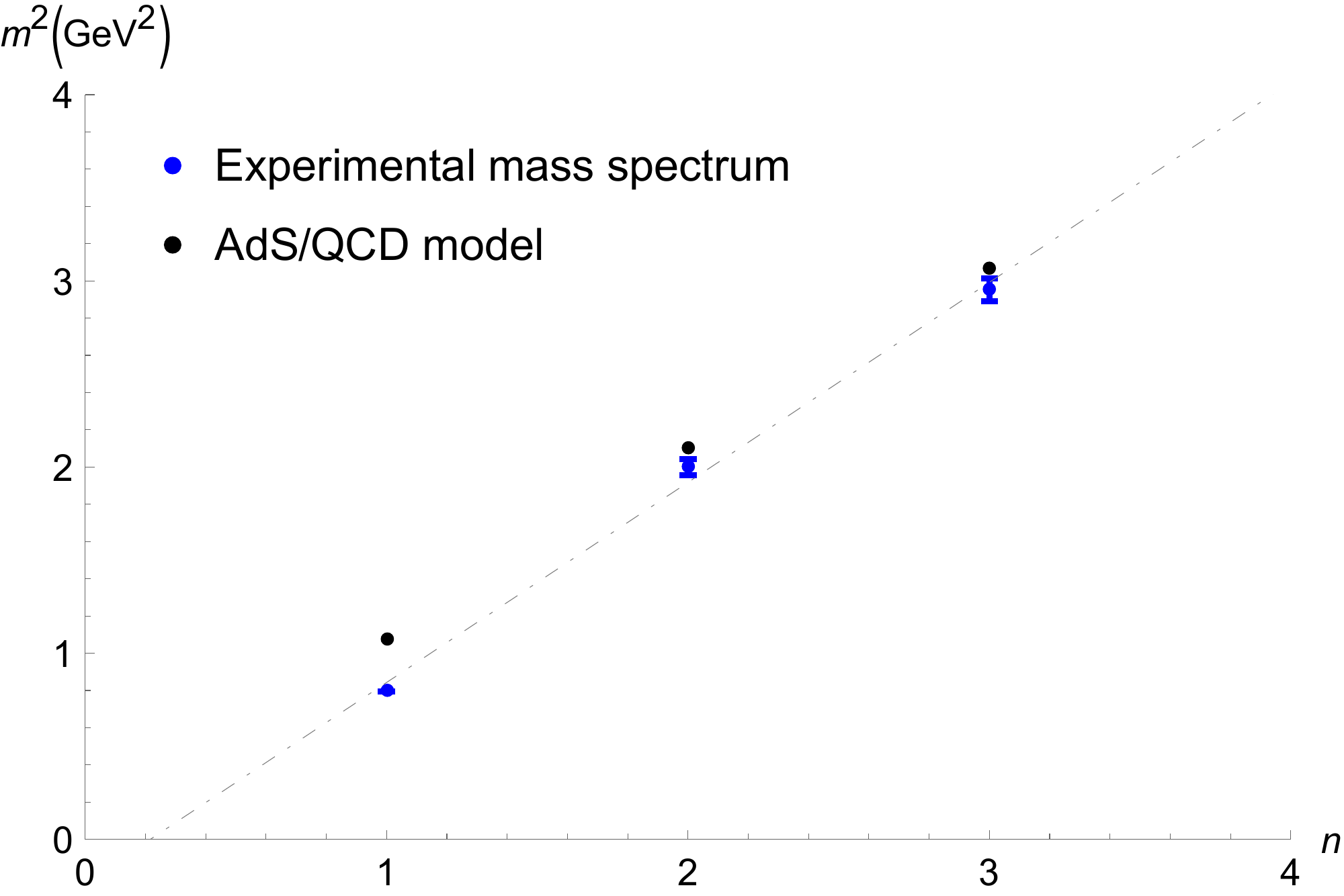}
	\caption{Mass spectrum of neutral strange vector kaons radial resonances: experimental values and AdS/QCD prediction, for $n=1,2,3$, with error bars. }
	\label{cen1}
\end{figure}

\begin{table}[H]
\begin{center}\medbreak
\begin{tabular}{||c|c||c|c|c||}
\hline\hline
$n$ & State & $M_{\scalebox{.67}{\textsc{Experimental}}}$ (MeV/$c^2$)  & $M_{\scalebox{.67}{\textsc{AdS/QCD}}}$ (MeV/$c^2$) &\;RE (\%)\;\\
       \hline\hline
\hline
1 &\;$K^*(892)\;$ & $895.55\pm0.20$ & 1038.4  & 13.8 \\ \hline
2 &\;$K^*(1410)\;$ & $1414\pm15 $ & 1451.0 & 2.6 \\ \hline
3& \;$\;K^*(1680)$& $1718 \pm 18$       & 1754.5 &  2.1   \\\hline
\hline\hline
\end{tabular}
\caption{Mass spectrum of neutral strange vector kaons resonances: experimental values and AdS/QCD prediction. The masses in the fourth column were obtained as solutions of Eq. \eqref{schr}, with $\bar{m}$ = 413 MeV, ${\scalebox{.94}{$\upalpha$}}= 0.056$ and ${\scalebox{.94}{$\kappa$}} = 532.06$ MeV/$c^2$. The fifth column displays the relative error regarding the third and fourth columns. } \label{scalarmasses1}
\end{center}
\end{table}
In Table \ref{scalarmasses1},  the $n=1$ state represents the neutral strange vector kaon ${K}^*(892)^0=d\bar{s}$.
The  linear Regge trajectory is illustrated in Fig. \ref{cen1}, together with the mass spectrum of strange  vector kaon resonances predicted by AdS/QCD.
A linear Regge trajectory, relating the squared mass of strange vector kaons as a linear function of the $n$ quantum number, fits experimental data in PDG \cite{pdg} and can be written as 
\beq \label{747}
m_n^2 = 1.0747\, n-0.23186.\label{lrt}
\eeq
 The expression $m_n^2=4\kappa^2n$ reasonably holds and is a better approximation for higher values of $n$.
The experimental mass spectrum of strange vector kaons resonances, listed in Table \ref{scalarmasses1}, can be reproduced by Eq. (\ref{lrt}) within a 5.8\% root-mean-square deviation (RMSD). Including error bars, as in Fig. \ref{cen1}, the linear Regge trajectory (\ref{lrt}) thoroughly matches the experimental mass spectrum of strange vector kaons resonances. 

\begin{figure}[H]
	\centering
	\includegraphics[width=8.9cm]{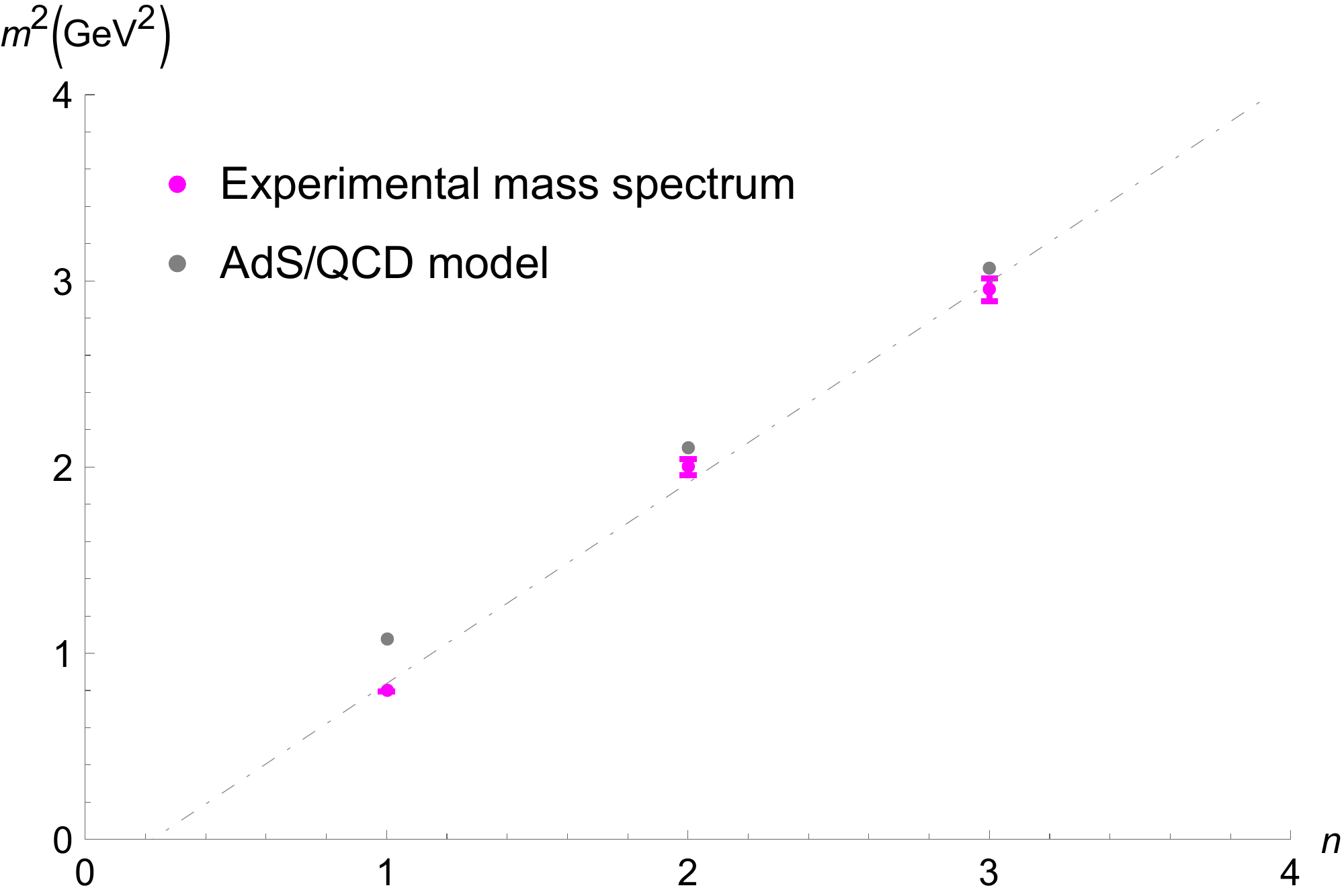}
	\caption{Mass spectrum of charged strange vector kaons radial resonances: experimental values and AdS/QCD prediction, for $n=1,2,3$. }
	\label{cen1k}
\end{figure}

Now there is another possibility, consisting of the strange vector mesons $K^*(892)^+=u\bar{s}$ and $K^*(892)^-=\bar{u} s$. It is illustrated in Table \ref{scalarmasses10}.
\begin{table}[H]
\begin{center}
\begin{tabular}{||c|c||c|c|c||}
\hline\hline
$n$ & State & $M_{\scalebox{.67}{\textsc{Experimental}}}$ (MeV/$c^2$)  & $M_{\scalebox{.67}{\textsc{AdS/QCD}}}$ (MeV/$c^2$) &\;RE (\%)\;\\
       \hline\hline
\hline
1 &\;$K^*(892)\;$ & $891.67\pm0.26$ & 1035.2  & 13.9 \\ \hline
2 &\;$K^*(1410)\;$ & $1414\pm15$ & 1446.9 & 2.3 \\ \hline
3& \;$\;K^*(1680)$& $1718 \pm 18$       & 1749.5  & 1.8  \\\hline
\hline\hline
\end{tabular}
\caption{Mass spectrum of  charged strange vector kaons radial resonances: experimental values and AdS/QCD prediction, for $n=1,2,3$. The masses in the fourth column, obtained by solving Eq. \eqref{schr}, with $\bar{m} = 411$ MeV/$c^2$, ${\scalebox{.94}{$\upalpha$}}= 0.0554$ and ${\scalebox{.94}{$\kappa$}} = 530.79$ MeV/$c^2$. The fifth column displays the relative error regarding the third and fourth columns. } \label{scalarmasses10}
\end{center}
\end{table}
In Table \ref{scalarmasses10} the $n=1$ state represents the possibilities $K^*(892)^+=u\bar{s}$ and $K^*(892)^-=\bar{u} s$. 
The  linear Regge trajectory is illustrated in Fig. \ref{cen1k}, together with the mass spectrum of vector kaon resonances predicted by AdS/QCD.
The linear Regge trajectory, relating the squared mass of vector kaons as a linear function of the $n$ quantum number, fits experimental data in PDG \cite{pdg}. The explicit expression for the linear Regge trajectory reads
\beq 
m_n^2 = 1.0782\, n-0.24112.\label{lrt1}
\eeq
The experimental mass spectrum of vector kaons resonances, listed in Table \ref{scalarmasses10}, can be reproduced by Eq. (\ref{lrt1}) within a 5.7\% RMSD. Including error bars, as in Fig. \ref{cen1k}, the linear Regge trajectory (\ref{lrt1}) thoroughly matches the experimental mass spectrum of vector kaons resonances.

\section{DCE of \smr}
\label{sec2}
The DCE is a useful tool for obtaining the mass spectrum of the next generation of \smre. To compute the DCE, one needs first to calculate the Fourier transform of the energy density. Indicating by ${\boldsymbol{q}}$ the spatial sector of the four-momentum $q$, one calculates  
\beq\label{fou}
{\scalebox{.89999}{$\tau_{00}$}}({\boldsymbol{q}}) = \frac{1}{(2\pi)^{k/2}}\int_{\scalebox{.72}{${\mathbb{R}^k}$}}\,{\scalebox{.89999}{$\tau_{00}$}}({\boldsymbol{r}})e^{-i{\boldsymbol{q}}\cdot {\boldsymbol{r}}}\,d^k {\boldsymbol{r}}.\eeq 
Thereafter, the  modal fraction  can be computed by the following expression   
\cite{Gleiser:2012tu,Gleiser:2011di,Gleiser:2018kbq}, 
\begin{eqnarray}
{\scalebox{.89999}{${\boldsymbol{\tau}}_{00}$}}({\boldsymbol{{q}}}) = \frac{\left|{\scalebox{.89999}{$\tau_{00}$}}({\boldsymbol{q}})\right|^{2}}{ \bigintsss_{\scalebox{.72}{${\mathbb{R}^k}$}}  \left|{\scalebox{.89999}{$\tau_{00}$}}({\bf{q}})\right|^{2}\,d^k{{\bf q}}}.\label{modalf}
\end{eqnarray} 
The DCE evaluates how much information is required to encode the energy density, also representing the weight of information carried by each momentum wave mode. The DCE is computed as 
\begin{eqnarray}
{\rm DCE}= - \int_{\scalebox{.72}{${\mathbb{R}^k}$}}\,\check{{\scalebox{.93}{${\boldsymbol{\tau}_{00}}$}}}({\boldsymbol{q}})\ln  \check{{\scalebox{.93}{${\boldsymbol{\tau}_{00}}$}}}({\boldsymbol{q}})\,d^k{{\bf q}},
\label{confige}
\end{eqnarray}
where {$\check{\scalebox{.89999}{${\boldsymbol{\tau}}_{00}$}}({\boldsymbol{q}})={\scalebox{.89999}{${\boldsymbol{\tau}}_{00}$}}({\boldsymbol{q}})/{\scalebox{.93}{${\boldsymbol{\tau}}^{\scalebox{.68}{\textsc{max}}}_{00}$}}({\boldsymbol{q}})$}, and ${\scalebox{.93}{${\boldsymbol{\tau}}^{\scalebox{.68}{\textsc{max}}}_{00}$}}({\boldsymbol{q}})$ denotes the maximum of 
{{\scalebox{.89999}{${\boldsymbol{\tau}}_{00}$}}. Besides, in the state ${\scalebox{.93}{${\boldsymbol{\tau}}^{\scalebox{.68}{\textsc{max}}}_{00}$}}({\boldsymbol{q}})$, the spectral density reaches its highest value. The unit of measurement of the DCE is nat/unit volume, where \emph{nat} stands for the  natural units of information  entropy. 1 bit equals $\ln 2$ nat. 
 The DCE represents scale information, as the power spectral density is  given by the Fourier transform of the 2-point correlator \cite{Gleiser:2018kbq}.

 The value $k=1$ is taken to evaluate the DCE of \smre, following the protocol in Eqs. (\ref{fou}) -- (\ref{confige}), since the AdS boundary has codimension one, concerning the bulk space in AdS/QCD. 
 Since the main ingredient to compute the DCE is the energy density, one
 first must switch the Lagrangian (\ref{lalag}) onto the expression 
 defining the energy density as the temporal component of the stress-energy-momentum,  
\beq
{\scalebox{.89999}{$\tau_{00}$}}\!&=&\!  \frac{2}{\sqrt{ -g }}\!\! \left[\frac{{\scalebox{.94}{$\partial$}} (\sqrt{-g}{{\scalebox{.94}{$\mathcal{L}$}}})}{{\scalebox{.94}{$\partial$}}{g^{00}}} \!-\!\frac{{\scalebox{.94}{$\partial$}}}{{\scalebox{.94}{$\partial$}}{ x^\beta }}  \frac{{\scalebox{.94}{$\partial$}} (\sqrt{-g} {{\scalebox{.94}{$\mathcal{L}$}}})}{{\scalebox{.94}{$\partial$}}\left(\frac{{\scalebox{.79}{$\,{\scalebox{.94}{$\partial$}}$}} g^{00}}{{\scalebox{.79}{$\,{\scalebox{.94}{$\partial$}}$}}x^\beta}\right)}
%  \!+\!\mathcal{T}^{mn}\!
  \right]
  \label{em1}
\nonumber\\\label{t001}
&=&\frac{1}{4\mathsf{g}_5^2}{\exp\left[{-({\scalebox{.94}{$\kappa$}} z)^{2-{\scalebox{.74}{$\upalpha$}}}}\right]}\,g_{00}F^{MN}F_{MN}^*-F^{0M}F_{0M}^*.
\eeq
 Regarding the plane wave solution  in the strange vector kaon rest frame,  $\mathsf{V}_\mu = \epsilon_\mu {\scalebox{.94}{$\upxi$}}(q, z)e^{-imc^2t}$, and  the transverse polarization vector  $\epsilon_\mu = (0, 1, 0, 0)^\intercal$ implies that  
\beq\label{t002}
\!\!\!\!\!{\scalebox{.89999}{$\tau_{00}$}}=\frac{z^2\exp\left[{-({\scalebox{.94}{$\kappa$}} z)^{2-{\scalebox{.64}{$\upalpha$}}}}\right]}{2\mathsf{g}_5^2R^2}\left[-\left({\scalebox{.94}{$\partial$}}_z{\scalebox{.94}{$\upxi$}}\right)^{2}+m_n^2{\scalebox{.94}{$\upxi$}}^2\right].
\eeq 
Once the energy density (\ref{t002}) has been  obtained, now it is straightforward to evaluate the Fourier transform of the energy density  by Eq. (\ref{fou}) and, subsequently, the modal fraction, and the DCE, respectively using Eqs. (\ref{modalf}, \ref{confige}).  
 This routine allows for assessing  the DCE  underlying the soft wall AdS/QCD. With it in hand, the mass spectrum of the next generation of \smr can be then obtained, by interpolating the experimental mass spectrum of the \smr in PDG \cite{pdg}. This approach, besides having more precision than solving Eq. (\ref{schr}) to obtain the mass spectrum, is phenomenologically  robust, since it relies on the experimental mass spectrum of \smr in PDG \cite{pdg}.  The DCE, using the protocol (\ref{fou}) -- (\ref{confige}) is numerically calculated. 
 We will first approach the  neutral \smr  in the next subsection, for which the first fundamental strange vector kaon state is the  $K^*(892)^0=~(d\bar{s})$.

 \subsection{DCE of neutral strange vector kaons}
 \label{ndce}
 The first analysis to be implemented will be the calculation of the DCE associated with the neutral  \smae. Namely, the first neutral \sma resonance, $n=1$,  represents the 
$K^*(892)^0 = d\bar{s}$, whereas the associated antikaon has structure $\bar{K}^*(892)^0=\bar{d} s$. For $n=2$, the state $K^*(1410)^0$ is described, whereas, for $n=3$, the state $K^*(1680)^0$ can be achieved, both with  the same constituent quark model. 
The masses in the fourth column were obtained as solutions of Eq. \eqref{schr}, with $\bar{m} = 413$ MeV/$c^2$, ${\scalebox{.94}{$\upalpha$}}= 0.056$ and ${\scalebox{.94}{$\kappa$}} = 532.06$ MeV/$c^2$.
The DCE of vector kaon resonances can be computed, using the energy density, as the temporal component of the energy-momentum tensor \eqref{t002}, in the DCE protocol, given by the subsequent use of Eqs. (\ref{fou}) -- (\ref{confige}). The DCE-based AdS/QCD hybrid method presented will be shown to consist of a more realistic complementary approach to standard bottom-up AdS/QCD predictions, as it takes the experimental masses of the first three vector kaon resonances to predict the mass spectrum of the next generation of strange vector kaons, which correspond to strange vector kaons resonances with higher excitation numbers.

The DCE for neutral strange vector kaons is exhibited in Table \ref{scalarmasses50},  encompassing the $K^*(892)^0 = d\bar{s}$ state with $n=1$.
\begin{table}[H]
\begin{center}
\begin{tabular}{||c|c|c||}
\hline\hline
$n$ & \;Neutral strange vector kaon resonance\; & DCE (nat) \\
       \hline\hline
\hline
\;1\; &\;$K^*(892)^0\;$ & 104.79   \\ \hline
\;2\; &\;$K^*(1410)^0\;$ & 176.05   \\ \hline
\;3\;& \;$\;K^*(1680)^0$& 253.21          \\\hline
\hline
\end{tabular}
\caption{DCE of the radial resonances in the neutral strange vector kaon family.} \label{scalarmasses50}
\end{center}
\end{table}
 Since the energy density is localized, the integration to compute the DCE (\ref{confige}) over the real line has been implemented within the limits of integration $|q|\leq 10^7$. In fact, implementing the integration for the DCE (\ref{confige}) in the interval $|q|\leq 10^{6}$ yields a difference below $10^{-9}$ in the DCE. This approximation error is lower than any experimental,  phenomenological, and numerical precision used heretofore. 
The same limits of integration  
 have been also used to compute the modal fraction \eqref{modalf}.
It can be seen by the plot of the modal fraction in Fig. \ref{mfs} that for $|q|\gtrsim 6\times10^{4},$ the modal fraction evanesces, sharply diminishing to zero, irrespectively of the value of $n$.
\begin{figure}[H]
	\centering
	\includegraphics[width=7cm]{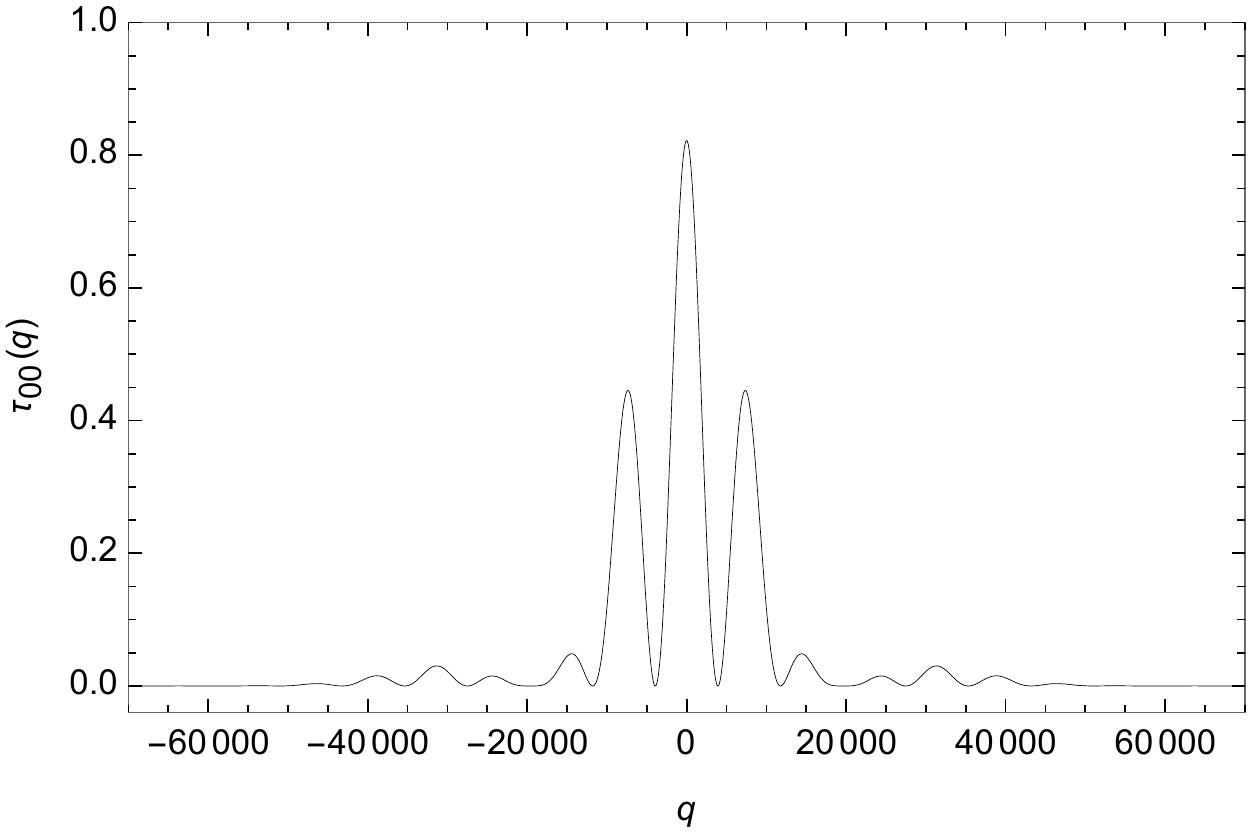}\qquad\quad
	\includegraphics[width=7cm]{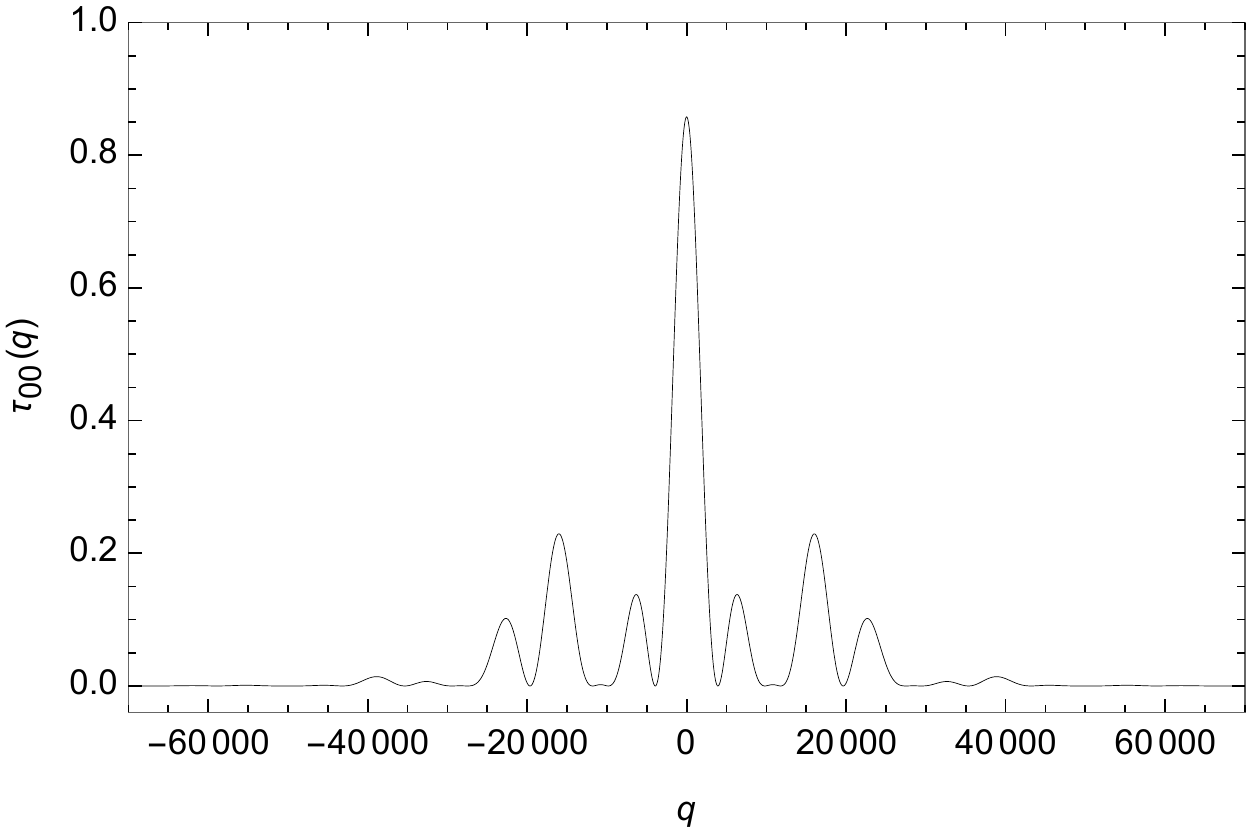}
	\includegraphics[width=7cm]{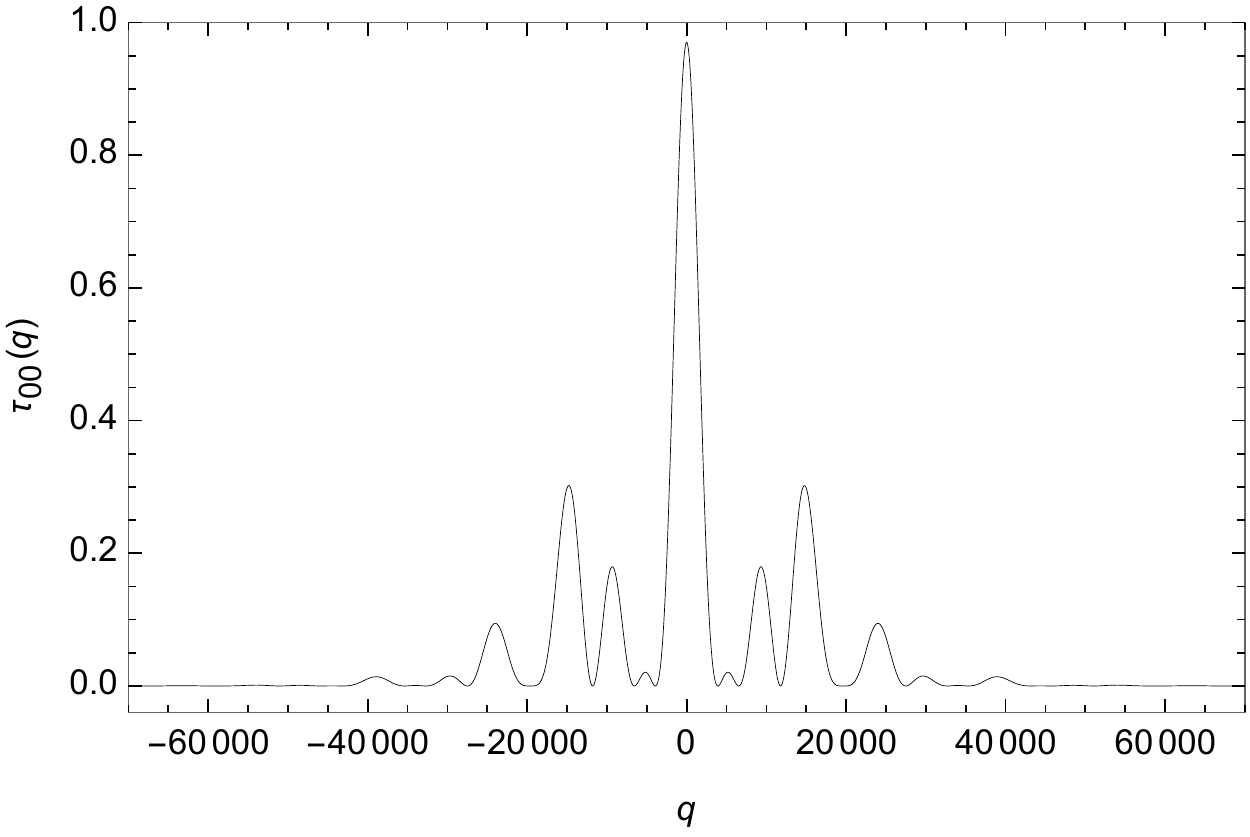}
	\caption{Modal fraction (\ref{modalf}) of the neutral vector kaon meson family, for  $n=1,2,3$, respectively corresponding to the $K^*(892)^0$, $K^*(1410)^0$, and $K^*(1680)^0$ states in PDG \cite{pdg}, as a function of the momentum $q$. The top panel on the left regards $n=1$, the top panel on the right corresponds to $n=2$, and the panel on the bottom refers to $n=3$.}
	\label{mfs}
\end{figure}

The  first form of DCE-Regge-like trajectories regards the DCE as a function of the $n$ radial excitation level of neutral  \smre. Fig. \ref{cen1} shows the obtained outcomes, whose cubic polynomial  interpolation of data in Table \ref{scalarmasses50} generates the first type of DCE-Regge-like trajectory, 
\begin{eqnarray}\label{itp1}
\!\!\!\!\!\!\!\!\!\!\!\!\clt{{\rm DCE}}_{K^{*}}&=& 1.1702 n^3 + 9.9717 n^2 + 49.5369n+ 46.4517.  \end{eqnarray}
 \clt{within $0.001\%$ RMSD}.

\begin{figure}[H]
	\centering
	\includegraphics[width=8.9cm]{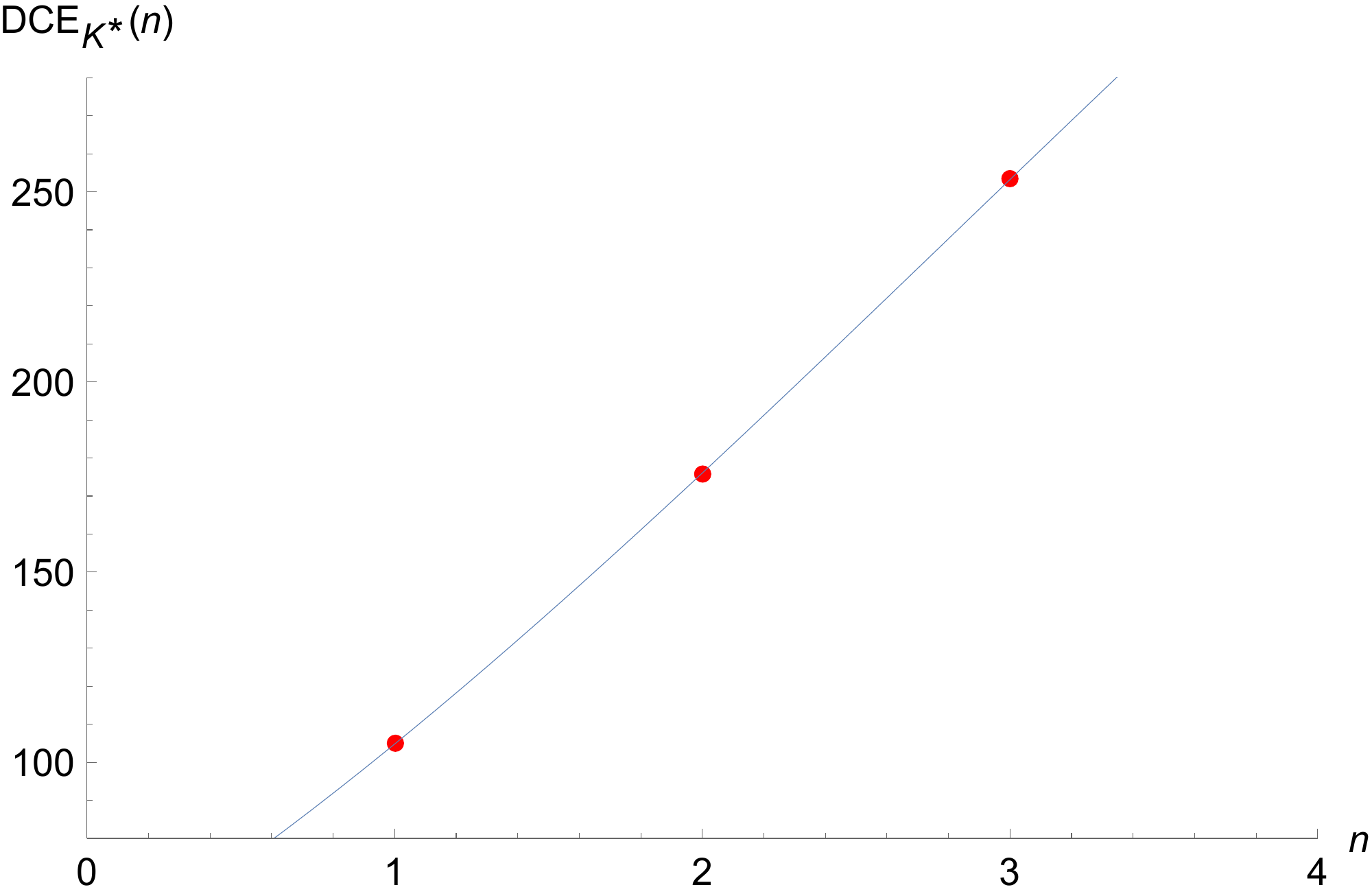}
	\caption{DCE of the neutral strange vector kaon family as a function of  the $n$ radial excitation level, for  $n=1,2,3$ (respectively corresponding to the $K^*(892)^0$, $K^*(1410)^0$, and $K^*(1680)^0$  resonances in PDG \cite{pdg}).  
The first form  of DCE-Regge-like trajectory (\ref{itp1}) is plotted as a dark gray line.}
	\label{cen1d}
\end{figure}

%Linear Regge trajectories illustrate the proportionality between the radial excitation level of light-flavor mesons and the square of their mass spectrum. However, it does not hold necessarily, when strange quarks enter the constituent model, as it must be necessary to describe vector kaons in this context. 

Now the DCE of the neutral strange vector kaon family can be  also realized  as a function of the experimental mass spectrum  of the strange vector kaons. In this way, the second type of DCE-Regge-like trajectory regards the experimentally detected  mass spectrum of strange vector kaon radial resonances \cite{pdg}. 
Having the DCE of all vector kaon radial resonances in Table \ref{scalarmasses50}, we can also plot the DCE as a function of the mass of each resonance, available in Table \ref{scalarmasses1}. The results are shown in Fig. \ref{cem11}, whose interpolation method generates the second type of DCE Regge trajectories in Eq. (\ref{itq11}). 
\begin{figure}[H]
	\centering
	\includegraphics[width=8.5cm]{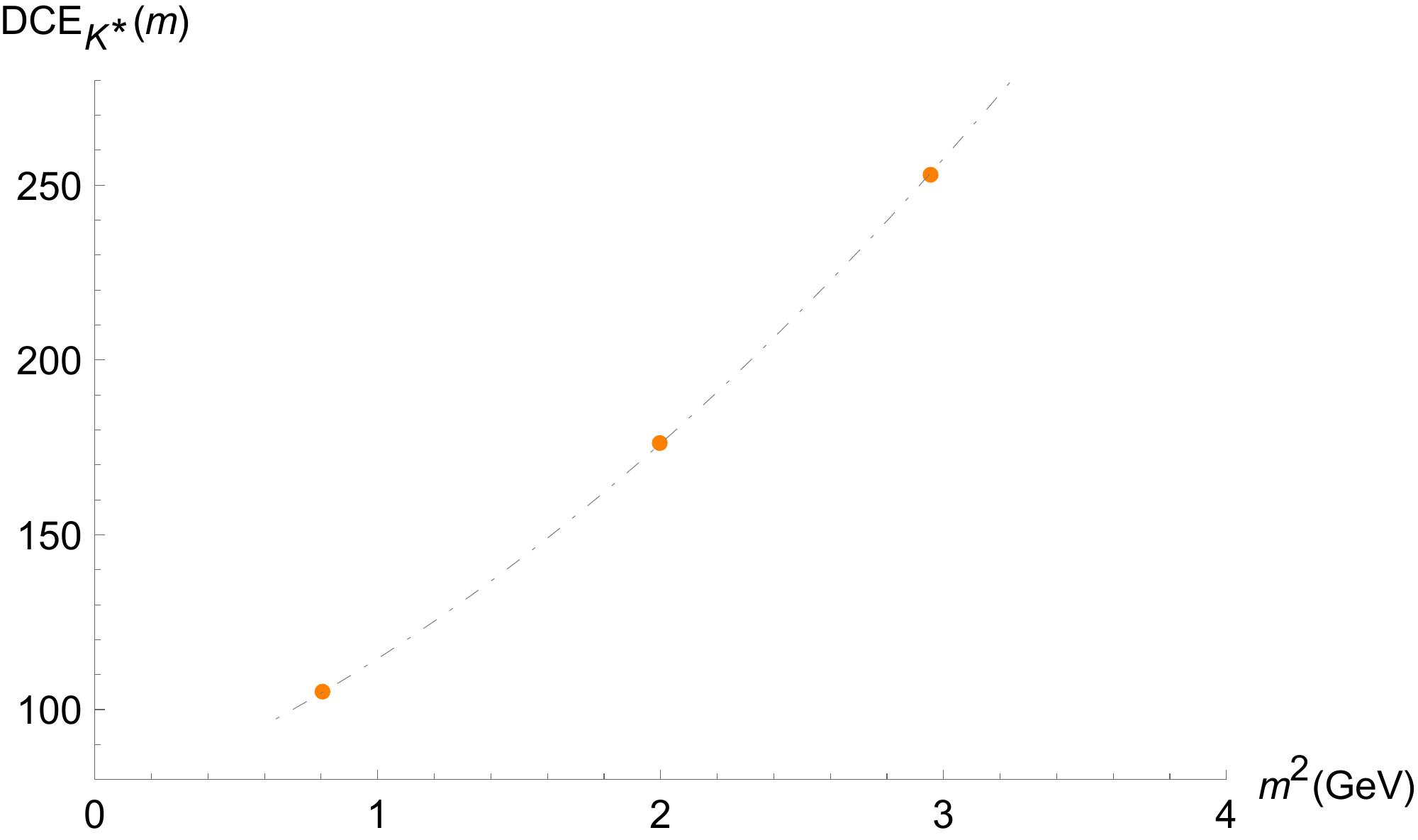}
	\caption{DCE of the neutral strange vector kaon  family  as a function of their squared mass, for  $n=1,2,3$ (respectively corresponding to the $K^*(892)^0$, $K^*(1410)^0$, and $K^*(1680)^0$  resonances in PDG \cite{pdg}). 
The DCE-Regge-like trajectory (\ref{itq11}) is represented by the interpolating dashed line.}
	\label{cem11}
\end{figure}
\noindent The second type of DCE-Regge-like trajectories, relating the DCE of the vector kaons to their squared mass spectra, $m^2$ (GeV${}^2/c^4$) read
\begin{eqnarray}
\label{itq11}
\!\!\!\!\!\!\!\!\!\!\!\!\!\!\!\!\!\!{\rm DCE}_{K^*(m)} \!&\!=\!& -0.143354 m^8+0.974149 m^6+7.73954
   m^4+33.6125 m^2+72.4105, 
   \end{eqnarray} within $0.001\%$  RMSD.    
   
Eqs. (\ref{itp1}) and (\ref{itq11}) carry the core of properties of the neutral strange vector kaon family. Besides, from Eq. (\ref{itp1}) one can promptly infer the DCE of elements in each family for higher values of the $n$ excitation level of resonances. Subsequently, with the DCE in hands for each $n$, one can substitute it on the left-hand side of Eq. (\ref{itq11}) and solve it,  deriving the mass of each \smr corresponding to $n>3$. In this way, the mass spectrum of new \smr  can be obtained. Besides, this method is based solely on the DCE and the experimental mass spectrum of the \smae, which is a more realistic technique when compared to the soft wall AdS/QCD, wherein the mass spectrum is read off the eigenvalues of the Schr\"odinger-like equation. Here we use the stress-energy-momentum  tensor of AdS/QCD to compute the DCE according to (\ref{fou}) -- (\ref{confige}), however, Eq.  (\ref{itq11}) is obtained by interpolation of the experimental mass spectrum of the neutral \smae, depicted in Fig. \ref{cem11}.

The first goal is to obtain the mass of the  $K^*_4$ strange vector kaon resonance. For it, the $n=4$ quantum number must be replaced into Eq. (\ref{itp1}), yielding the corresponding DCE to be equal to \clt{329.248 nat}. Then, substituting it directly into the left-hand side of Eq. (\ref{itq11}), one can obtain the  algebraic solution for the eighth-degree polynomial in the $m$ variable. This procedure yields the mass  \clt{$m_{K^*_4}= 1933.81$ MeV/$c^2$}. This method can be also applied to higher excitations  in the neutral strange vector kaon family. When $n=5$ is inserted in Eq. (\ref{itp1}), the DCE associated with the $K^*_5$ state is equal to \clt{397.143 nat}. Hence, one can subsequently  work out Eq. (\ref{itq11}) for this value. It yields the $K^*_5$ vector kaon resonance mass value  \clt{$m_{K^*_5}= 2092.71$ MeV/$c^2$}.  
Correspondingly,  Eq. (\ref{itp1}) can be now evaluated for the $n=6$ vector kaon resonance, yielding the DCE of the $K^*_6$ excitation to assume the value \clt{449.873 nat}. Therefore Eq. (\ref{itq11}) is solved for this value, yielding the mass  \clt{$m_{K^*_6}= 2203.43$ MeV/$c^2$}. These results are compiled in Table 
\ref{scalarmasses102}. 
	\begin{table}[H]
\begin{center}\begin{tabular}{||c|c||c|c||}
\hline\hline
$n$ & State & $M_{\scalebox{.67}{\textsc{Experimental}}}$ (MeV/$c^2$)  & $M_{\scalebox{.67}{\textsc{AdS/QCD and hybrid}}}$ (MeV/$c^2$) \\
       \hline\hline
\hline
1 &\;$K^*(892)\;$ & $891.67\pm0.26$ & 1035.2   \\ \hline
2 &\;$K^*(1410)\;$ & $1414\pm15 $ & 1446.9   \\ \hline
3& \;$\;K^*(1680)$& $1718 \pm 18$       & 1749.5     \\\hline
4& \;$K^*_4\;$& ---------  & 1933.81${}^\star$  \\\hline
5& \;$K^*_5\;$& ---------     & 2092.71${}^\star$   \\\hline
6& \;$K^*_6\;$&  ---------   & 2203.43${}^\star$ \\\hline
\hline\hline
\end{tabular}
\caption{Table \ref{scalarmasses1} completed with the higher $n$  resonances  of the neutral strange vector kaon meson family. The extrapolated masses for $n=4, 5, 6$ in the fourth column indicated with a `` ${}^\star$ '' denote the values extrapolated by the concomitant use of the DCE-Regge-like trajectories (\ref{itp1}, \ref{itq11}), interpolating the experimental masses for $n=1, 2, 3$. } \label{scalarmasses102}
\end{center}
\end{table}
The $K^*_5$ strange vector kaon resonance, whose mass is  predicted to be 2092.71 MeV/$c^2$, might match the  $X(2075)$ further mesonic state in \cite{pdg}, which has experimental mass  $2075\pm12\pm5$ MeV/$c^2$.
It might also match the  $X(2080)$ further mesonic state in \cite{pdg}, which has experimental mass  $2080\pm10$ MeV/$c^2$. In addition, the $K^*_6$ strange vector kaon resonance, whose mass is  predicted to be 2203.43  MeV/$c^2$, might match the  $X(2210)$ further mesonic state in \cite{pdg}, which has experimental mass  $2210^{+79}_{-15}$ MeV/$c^2$ \cite{pdg}. 
%But it is just a speculation, in the sense that the data in PDG does not forbid these identifications. 

 \subsection{DCE of charged strange vector kaons}
 \label{cdce}
 Now charged strange vector kaons will be addressed. The first configuration regards the state 
 $K^*(892)^+=(u\bar{s})$, whereas the another possibility is given by 
$K^*(892)^-~=~(\bar{u} s)$, both for $n=1$.
Correspondingly, the same constituent quark models are used to describe the $K^*(1410)^\pm$ and $K^*(1680)^\pm$ charged strange vector kaons. 
 Whatever the case, the average quark mass reads 
 with $\bar{m} = 411$ MeV/$c^2$, and the deformation parameter in the dilaton is given by ${\scalebox{.94}{$\upalpha$}}= 0.0554$, whereas  ${\scalebox{.94}{$\kappa$}} = 530.79$ MeV/$c^2$. The DCE of strange vector kaon resonances can be computed, using the energy density, as the temporal component of the energy-momentum tensor \eqref{t002}, in the DCE protocol given by the subsequent numerical calculations in Eqs. (\ref{fou}) -- (\ref{confige}). 

\begin{table}[H]
\begin{center}
\begin{tabular}{||c|c|c||}
\hline\hline
$n$ & State & DCE (nat) \\
       \hline\hline
\hline
1 &\;$K^*(892)\;$ & 106.98   \\ \hline
2 &\;$K^*(1410)\;$ & 179.11   \\ \hline
3& \;$\;K^*(1680)$& 258.15          \\\hline
\hline\hline
\end{tabular}
\caption{DCE of the radial resonances in the charged strange vector kaons  family, for the $u\bar{s}$ and  $\bar{u}s$ quark constituent model.} \label{scalarmasses60}
\end{center}
\end{table}

The  first type of DCE-Regge-like trajectories takes into account the DCE of charged strange vector kaons as a function of the quantum number $n$. Fig. \ref{cen11} displays the values of the DCE for $n=1,2,3$ and also the interpolation of data in Table \ref{scalarmasses60}, whose expression is given by 
\begin{eqnarray}\label{itp2}
\!\!\!\!\!\!\!\!\!\!\!\!\clt{{\rm DCE}_{K^*}}\!\!&\!=\!&-1.09872  n^3 + 10.0473n^2 + 49.679n+48.3523.  \end{eqnarray}
 \clt{within $0.001\%$ RMSD}. 
   
\begin{figure}[H]
	\centering
	\includegraphics[width=8.9cm]{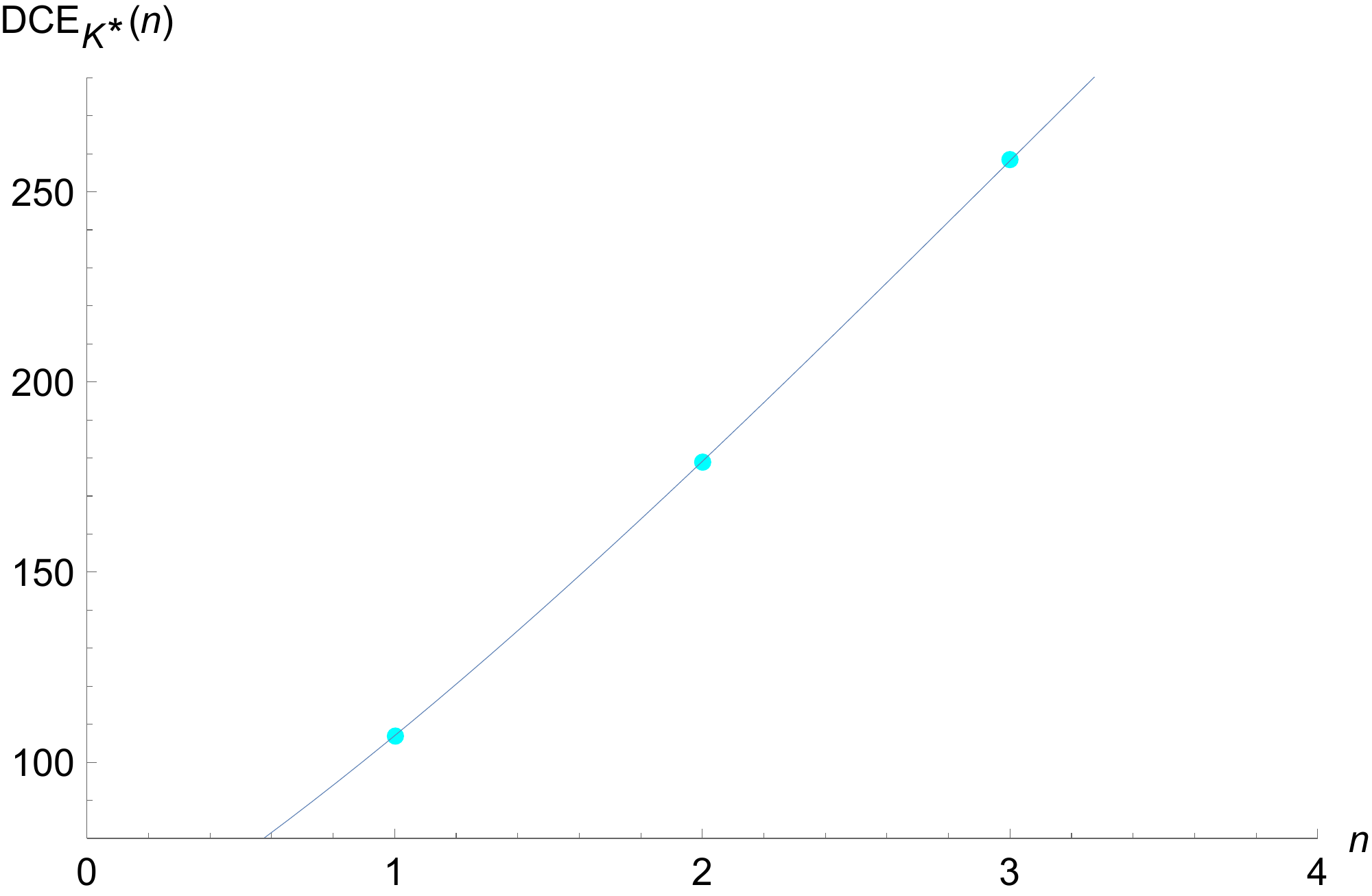}
	\caption{DCE of the vector kaon meson family as a function of the 
  $n$ radial excitation level, for  $n=1,2,3$ (respectively corresponding to the $K^*(892)^\pm$, $K^*(1410)^\pm$, and $K^*(1680)^\pm$  resonances in PDG \cite{pdg}).  
The first form  of DCE-Regge-like trajectory (\ref{itp2}) is plotted as a dark gray line.}
	\label{cen11}
\end{figure}
The DCE of \sma can be  also realized  as a function of their mass spectrum. In this way, the second type of DCE-Regge-like trajectory regards the   mass spectrum  of the \sma experimentally detected and reported in PDG \cite{pdg}. 
Having the DCE of all excited states in the strange vector kaon family, the DCE can be also plotted as a function of the mass of each resonance, which can be read off Table \ref{scalarmasses10}. The results are shown in Fig. \ref{cem22} whose interpolation method generates the second type of DCE Regge trajectories in Eq. (\ref{itq111}). These equations are the second type of DCE Regge trajectories.

\begin{figure}[H]
	\centering
	\includegraphics[width=8.5cm]{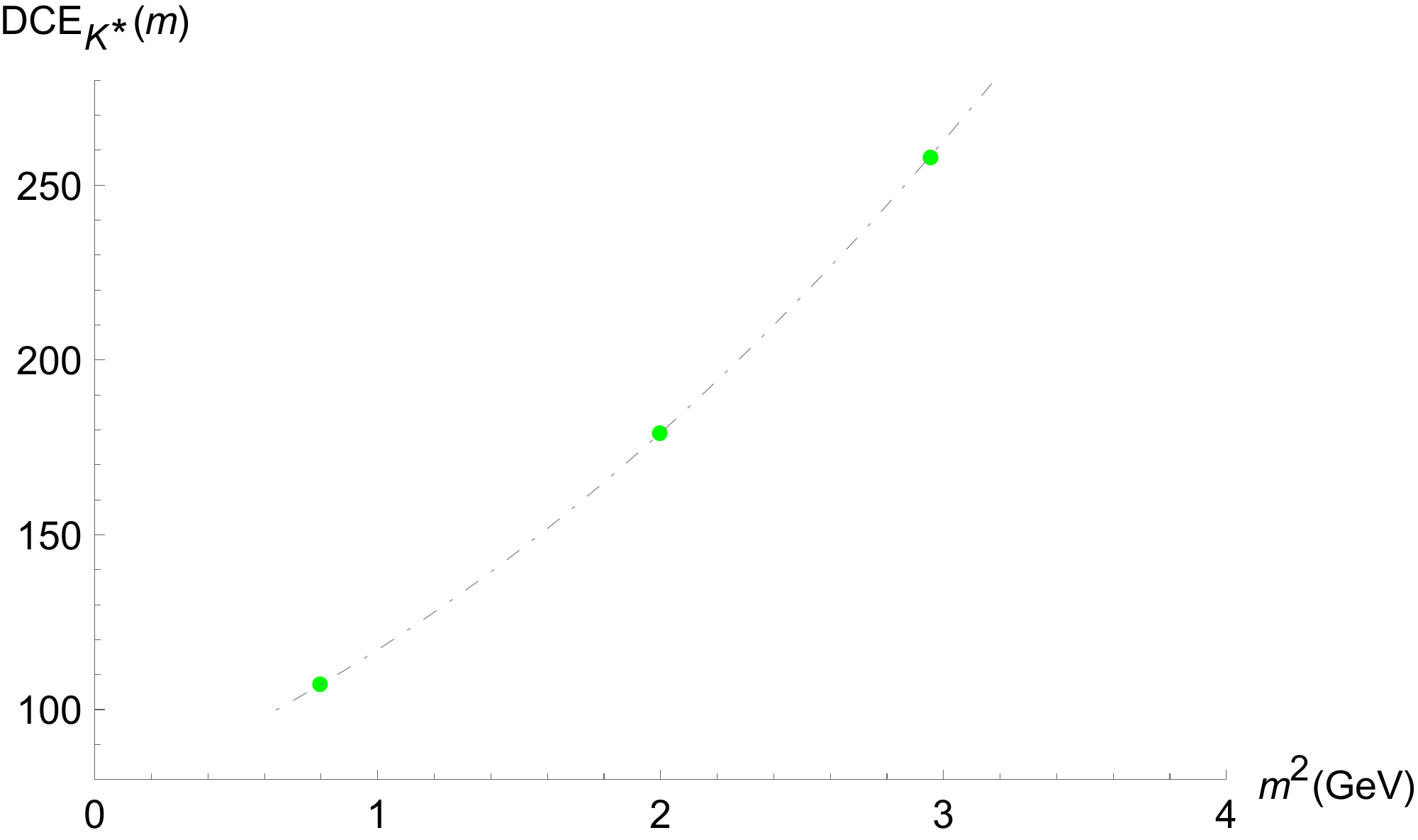}
	\caption{DCE of the charged \smre, for  $n=1,2,3$, as a function of their squared mass  (respectively corresponding to the $K^*(892)^\pm$, $K^*(1410)^\pm$, and $K^*(1680)^\pm$  resonances in PDG \cite{pdg}). 
The DCE-Regge-like trajectory (\ref{itq111}) is represented by the interpolating dashed line.}
	\label{cem22}
\end{figure}
\noindent The second type of DCE-Regge-like trajectory, which makes the DCE of the charged strange vector kaons to be a function of their squared mass spectra, $m^2$ (GeV${}^2/c^4$), has the following expression: 
\begin{eqnarray}
\!\!\!\!\!\!\!\!\!\!\!\!\!\!\!\!\!\!\clt{{\rm DCE}_{K^*}(m)} \!&\!=\!& -0.118751m^8+1.00297m^6+7.7114
   m^4+33.6419 m^2+74.9008, \label{itq111}
   \end{eqnarray} \clt{within $0.1\%$  RMSD}.    
   
From Eqs. (\ref{itp2}, \ref{itq111}), several features of charged strange vector kaons can be analyzed. Looking at Eq. (\ref{itp2}) one can promptly infer the DCE of elements in each family for higher values of the $n$. Subsequently, one can replace each value of $n$ on the left-hand side of Eq. (\ref{itq111}) and solve it. With this method, one can find the mass of charged \smr for $n>3$.  We want to determine the mass of the  $K^*_4$ charged \smrs, corresponding to $n=4$. 
When  putting $n=4$ back into Eq. (\ref{itp2}), the DCE amounts to \clt{337.507 nat}. Afterward, using this value of DCE on the left-hand side of Eq. (\ref{itq111}), one can obtain the  algebraic solution for the $m$ variable, yielding the mass of the  $K^*_4$ resonance to be  \clt{$m_{K^*_4}= 1933.16$ MeV/$c^2$}. Now, accomplishing the same technique to the $n=5$ excitation level in Eq. (\ref{itp2}), the DCE underlying the $K^*_5$ resonance reads \clt{410.590 nat}. Subsequently  working out Eq. (\ref{itq111}) for this value of the DCE implies the $K^*_5$ resonance mass to assume the value  \clt{$m_{K^*_5}= 2118.11$ MeV/$c^2$}.  
Analogously,  the $n=6$ excitation level can be now implemented into Eq. (\ref{itp2}), implying the DCE of the $K^*_6$ resonance to be equal to \clt{470.807 nat}. When Eq. (\ref{itq111}) is then resolved, using this value of DCE, the $K^*_6$ resonance mass  \clt{$m_{K^*_6}= 2239.02$ MeV/$c^2$} is acquired. These results are compiled in Table 
\ref{scalarmasses100}. 
	\begin{table}[H]
\begin{center}\begin{tabular}{||c|c||c|c||}
\hline\hline
$n$ & State & $M_{\scalebox{.67}{\textsc{Experimental}}}$ (MeV/$c^2$)  & $M_{\scalebox{.67}{\textsc{Theory}}}$ (MeV/$c^2$) \\
       \hline\hline
\hline
1 &\;$K^*(892)\;$ & $891.67\pm0.26$ & 1035.2   \\ \hline
2 &\;$K^*(1410)\;$ & $1414\pm15 $ & 1446.9   \\ \hline
3& \;$\;K^*(1680)$& $1718 \pm 18$       & 1749.5     \\\hline
4& \;$K^*_4\;$& ----------  & 1933.16${}^\star$  \\\hline
5& \;$K^*_5\;$& -----------     & 2118.11${}^\star$   \\\hline
6& \;$K^*_6\;$&  -----------   & 2239.02${}^\star$ \\\hline
\hline\hline
\end{tabular}
\caption{Table \ref{scalarmasses10} completed with resonances $n=4, 5, 6$ (with an asterisk) of the charged strange vector kaon meson family. The extrapolated masses  in the fourth column indicated with a `` ${}^\star$ '' specify that they are the result of the concomitant use of DCE-Regge-like trajectories (\ref{itp2}, \ref{itq111}), interpolating the experimental masses from $n=1, 2, 3$ resonances. } \label{scalarmasses100}
\end{center}
\end{table}
The $K^*_6$ vector kaon resonance, whose mass is  predicted to be 2198.13 MeV/$c^2$, might match the  $X(2210)$ further mesonic state in \cite{pdg}, which has experimental mass  $2210^{+79}_{-15}$ MeV/$c^2$ \cite{pdg}. Comparing the predicted mass spectrum for neutral and charged strange vector kaons, respectively in Tables \ref{scalarmasses102} and \ref{scalarmasses100}}, 
a difference of 0.03\% for the neutral and charged $K^*_4$ resonances can be found, which rises to 1.20\% when comparing the difference between the neutral and charged $K^*_5$ resonance, whereas this difference arises to 1.59\%,  for the $K^*_6$ resonance. These small differences in the mass spectrum for neutral and charged \smr do not increase the chances of discovering new candidates in experimental data in PDG \cite{pdg}, but they offer new possibilities for future experimental runnings though.

\section{DCC of the strange vector kaon family}
\label{sec21}

Complementarily to the DCE, Ref. \cite{Gleiser:2018jpd} introduced another important configurational information measure, namely, the differential configurational complexity (DCC). For defining it, another kind of  modal fraction must be introduced, whose normalization takes into account the square of the contribution of the maximum wave mode to the energy density, instead of Eq. (\ref{modalf}). The Fourier transform (\ref{fou}) is the first step to calculate the DCC, followed by the modal fraction  
\begin{eqnarray}\label{mf2}
\mathsf{f}_{\scalebox{.93}{${{\tau}}_{00}$}}({\boldsymbol{q}}) = 
\frac{\left|{\scalebox{.93}{${{\tau}}_{00}$}}({\boldsymbol{q}})\right|^2}{\left|{\scalebox{.93}{${{\tau}}^{\scalebox{.68}{\textsc{max}}}_{00}$}}({\boldsymbol{q}})\right|^2}.
\end{eqnarray}
The respective contribution of each wave mode to the power spectral density can be still evaluated by the modal fraction (\ref{mf2}). For a uniform power spectral density, the intrinsic complexity is lower, while for a nonuniform power spectral density, the complexity increases.
The DCC is defined by  
\cite{Gleiser:2018jpd}, 
\begin{eqnarray}\label{dcc}
S_{\scalebox{.55}{\textsc{DCC}}} = - \int_{\mathbb{R}^k} \,\mathsf{f}_{\scalebox{.93}{${\boldsymbol{\tau}}_{00}$}}({\boldsymbol{q}})\ln\mathsf{f}_{\scalebox{.93}{${\boldsymbol{\tau}}_{00}$}}({\boldsymbol{q}})\,d^k{\boldsymbol{q}}.
\end{eqnarray}
The DCC \eqref{dcc} vanishes for wave modes with the same weight \cite{Gleiser:2018jpd}. Non-interacting noise is related to a uniform power spectral density, producing the largest DCE value and vanishing DCC, though a plane wave has vanishing DCE and DCC. It illustrates the relevant interpretation regarding the DCC as a configurational information  measure of the complexity of shape. 
The shape associated with a given physical structure can be expressed by the 2-point correlator, whose Fourier transform is the power spectral density, encoding shape into the momentum wave modes  \cite{Gleiser:2018jpd}. 

Again, for  computing the DCC of \smr the $k=1$ is chosen in Eqs. (\ref{fou}), (\ref{mf2}) and (\ref{dcc}). Replacing the Lagrangian (\ref{lalag}) into the stress-energy-momentum  tensor field expression \eqref{em1} and using the energy density \eqref{t002}, one can first compute its Fourier transform, the modal fraction, and the DCC, respectively using  Eqs. (\ref{fou}), (\ref{mf2}) and (\ref{dcc}). In this way, the mass spectrum of \smr can be studied, also in the context of the DCC. 

 \subsection{DCC of neutral strange vector kaons}
 \label{ndcc}
\begin{figure}[H]
	\centering
	\includegraphics[width=7.8cm]{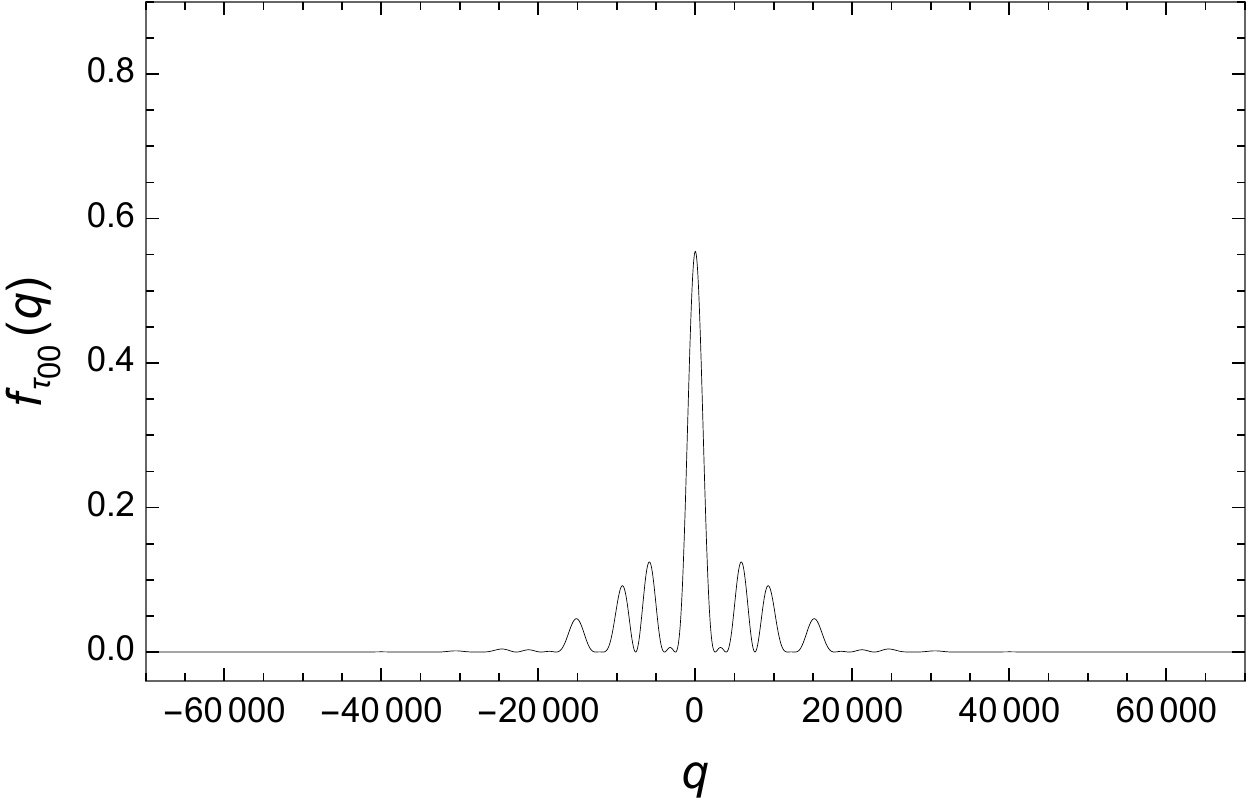}\qquad\quad
	\includegraphics[width=7.8cm]{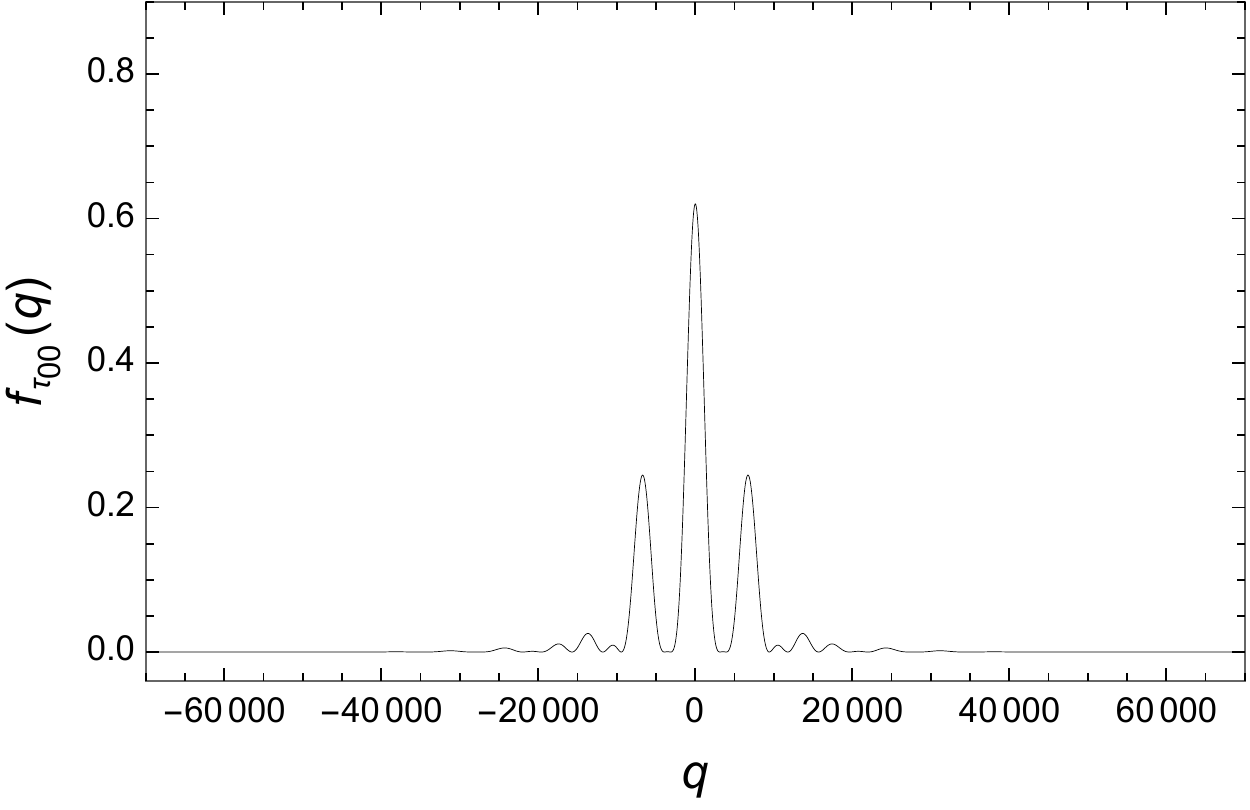}
	\includegraphics[width=7.8cm]{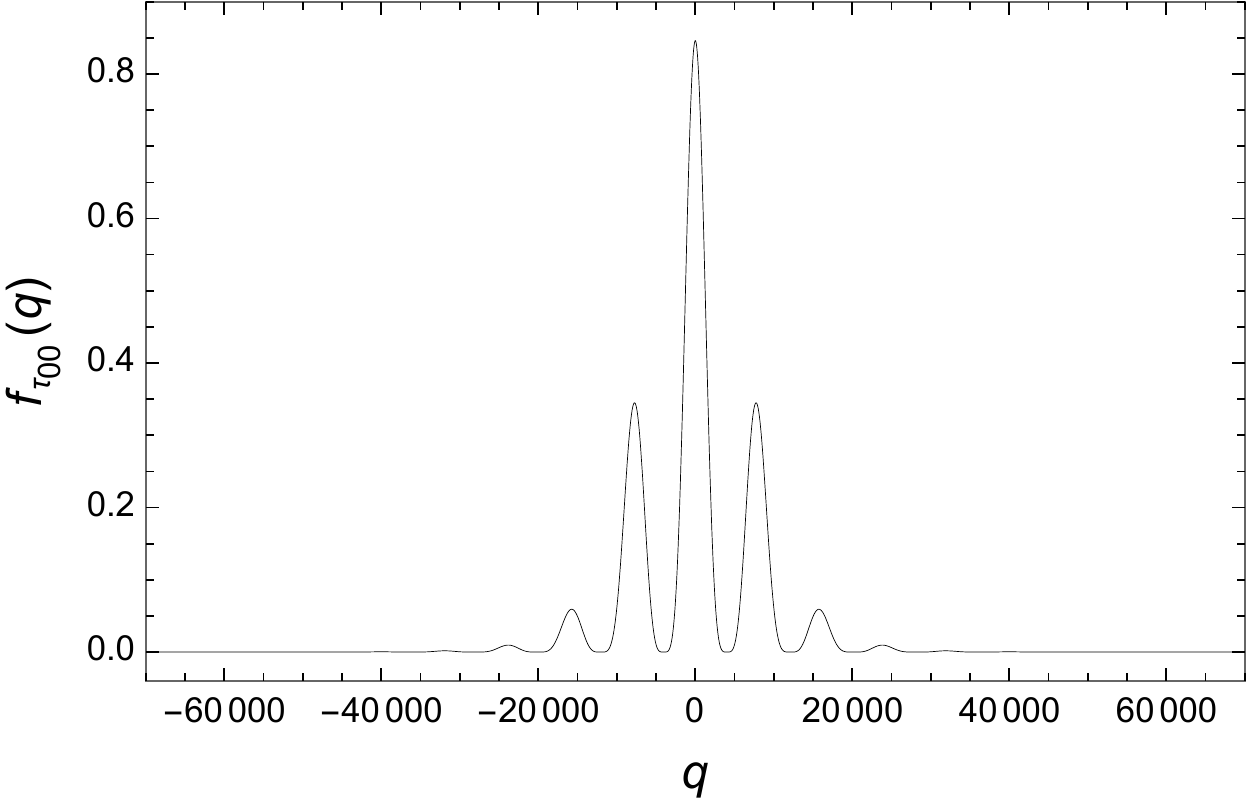}
	\caption{Modal fraction (\ref{mf2}) for the DCC of the neutral strange vector kaon meson family as a function of the momentum $q$, for  $n=1,2,3$, respectively corresponding to the $K^*(892)^0$, $K^*(1410)^0$, and $K^*(1680)^0$ states in PDG \cite{pdg}. The top panel on the left regards $n=1$, the top panel on the right corresponds to $n=2$ and the panel on the bottom refers to $n=3$.}
	\label{mfs1}
\end{figure}

The DCC can be also evaluated for neutral \smae. Namely, we can  take the $d\bar{s}$ strange vector kaons.  In this case, the average quark mass reads  $\bar{m} = 413$ MeV/$c^2$, 
 and the parameter that is responsible for the dilaton deformation is given by  ${\scalebox{.94}{$\upalpha$}}= 0.056$, whereas  ${\scalebox{.94}{$\kappa$}} = 532.06$ MeV/$c^2$. 
The corresponding DCC is numerically computed and  exhibited in Table \ref{scalarmasses5}. 
\begin{table}[H]
\begin{center}
\begin{tabular}{||c|c|c||}
\hline\hline
$n$ & State & DCC  \\
       \hline\hline
\hline
1 \;&\;$K^*(892)\;$ & \;279.12 \;  \\ \hline
2\; &\;$K^*(1410)\;$ & \;510.92\;   \\ \hline
3\;& \;$\;K^*(1680)\;$& \;777.20\;    \\\hline
\hline\hline
\end{tabular}
\caption{DCC of the neutral strange vector meson radial resonances.} \label{scalarmasses5}
\end{center}
\end{table}
Similarly to the analysis of the DCE of \smae, due to the energy density localization, here also the integration to compute the DCC (\ref{dcc}) over the real line, has been numerically evaluated within the range of integration $|q|\leq \times 10^6$. Implementing the integration for the DCC (\ref{dcc}) in the interval $|q|\leq 10^{7}$ produces a difference below $10^{-8}$ in the DCC, when compared to considering the interval $|q|\lesssim  10^8$. A variation of $10^{-8}$ is lower than any experimental,  phenomenological, and numerical precision used up to now. Besides, one can see in Fig. \ref{mfs1} that for $|q|\gtrsim 4\times10^{4},$ the modal fraction evanesces, bluntly fading away and fastly approaching zero, irrespectively of the value of $n$.

\begin{figure}[H]
	\centering
	\includegraphics[width=8.9cm]{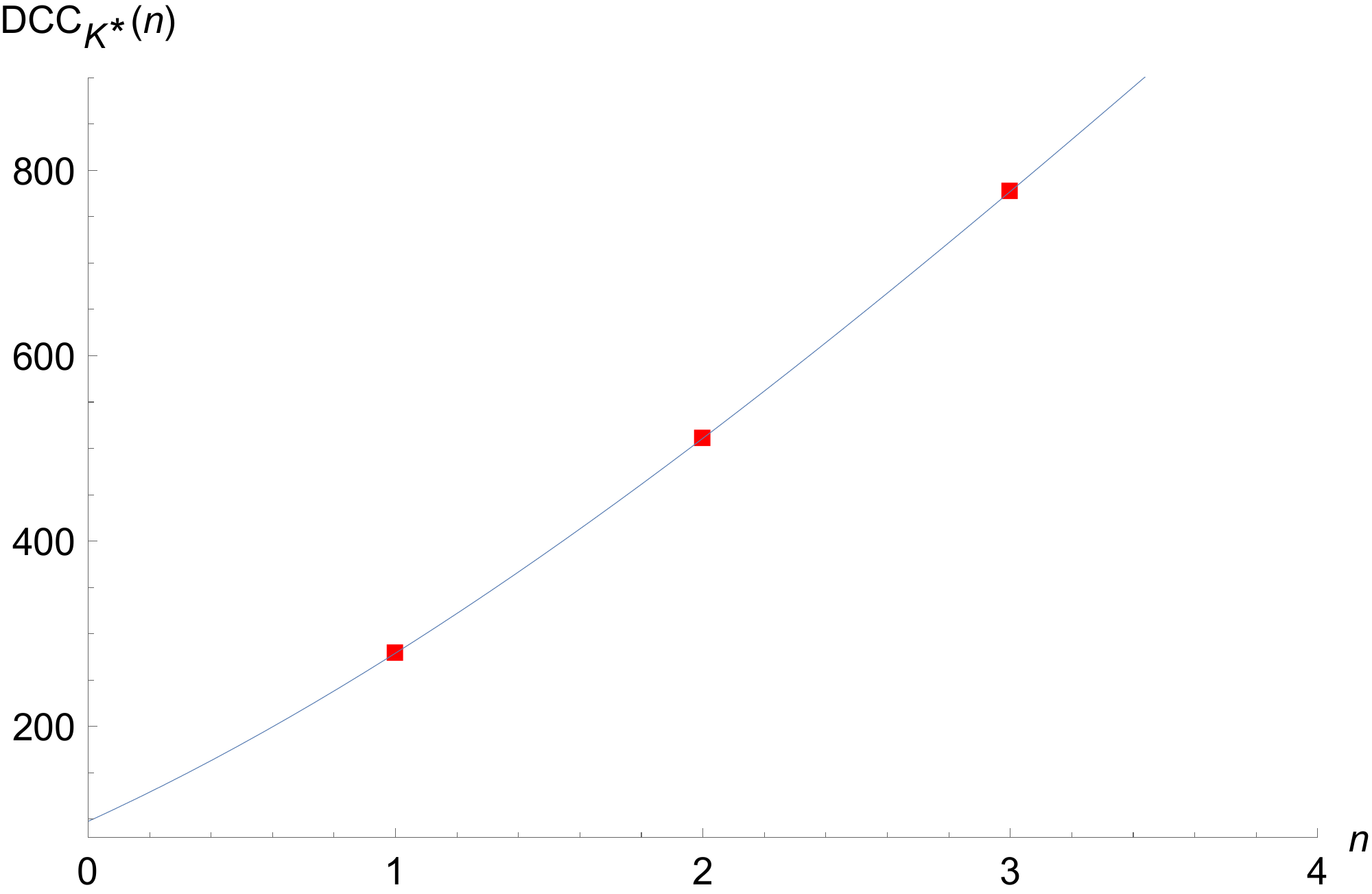}
	\caption{DCC of the neutral strange vector kaon meson family as a function of $n$, for  $n=1,2,3$ (respectively corresponding to the $K^*(892)^0$, $K^*(1410)^0$, and $K^*(1680)^0$ resonances in PDG \cite{pdg}).  
The first form  of DCC-Regge-like trajectory (\ref{itp2}) is plotted as a dark gray line.}
	\label{cen3}
\end{figure}

\begin{figure}[H]
	\centering
	\includegraphics[width=8.5cm]{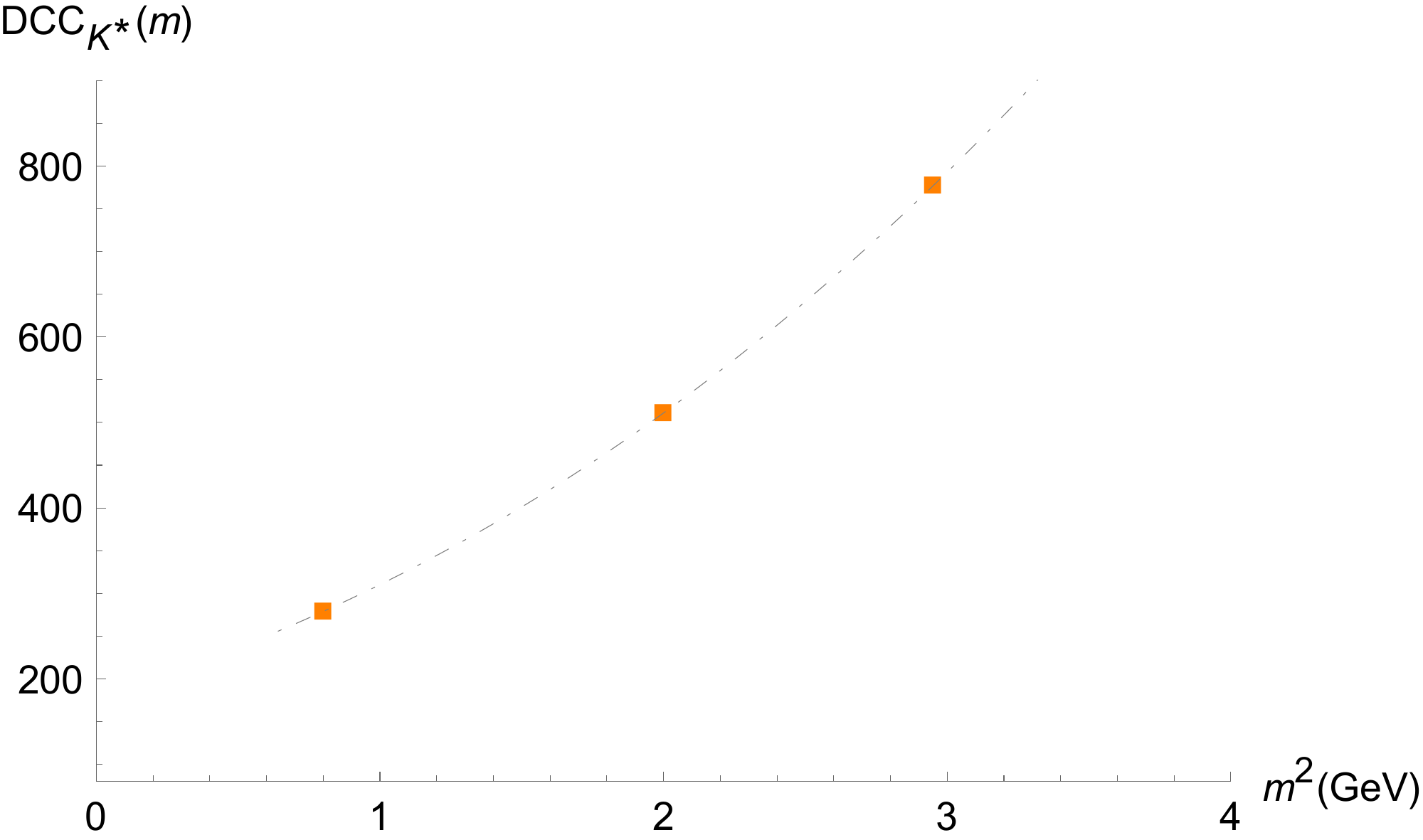}
	\caption{DCC of the neutral strange vector meson family as a function of their squared mass, for  $n=1,2,3$ (respectively corresponding to the $K^*(892)^0$, $K^*(1410)^0$, and $K^*(1680)^0$ resonances in PDG \cite{pdg}). 
The DCC-Regge-like trajectory is represented by the interpolating dashed line.}
	\label{cem44}
\end{figure}

The  first form of DCC-Regge-like trajectory is simply the graph of 
the  DCC as a function of the $n$ quantum number of \smre, illustrated by  Fig. \ref{cen3}.
Polynomial  interpolation of data in Table \ref{scalarmasses5} generates  the first kind of DCC-Regge-like trajectories, 
\begin{eqnarray}\label{itp3}
\!\!\!\!\!\!\!\!\!\!\!\!\clt{{\rm DCC}}\!\!&\!=\!& -2.63272 n^3+33.0363 n^2+151.120 n+97.5963,  \end{eqnarray}
within $0.001\%$ RMSD. 
The DCC of \smr can be also taken into account as a function of their mass spectrum, yielding a second kind of DCC-Regge-like trajectories. For it, once the DCC has been computed (Table \ref{scalarmasses5}), the DCC can be plotted as a function of the experimental mass spectrum, available in Table \ref{scalarmasses1}. The related results are plotted in Fig. \ref{cem44}.
\noindent  Interpolation of data engenders the second kind of DCC Regge trajectory for neutral strange vector kaon 
resonances, relating the DCC  to the squared mass spectra, $m^2$ (GeV${}^2/c^4$):
\begin{eqnarray}
\label{itq44}
\!\!\!\!\!\!\!\!\!\!\!\!\!\!\!\!\!\!{\rm DCC}_{K^*} \!&\!=\!& -0.283662 m^8+3.64317 m^6+25.6752
   m^4+102.600 m^2+178.557, \label{itq33}
   \end{eqnarray} \clt{within $0.001\%$  RMSD}.    
For each $n>3$, one can substitute the value of the DCC on the left-hand side of Eq. (\ref{itq44}) and algebraically solve it for the variable $m$. Using this method, the mass spectrum of the next generation in the neutral \sma family can be estimated, for $n=4,5,6,\ldots$. This approach is fixed on the experimental mass spectrum of the neutral strange vector kaon family, complementing the AdS/QCD with a configurational information measure. The stress-energy-momentum  tensor of AdS/QCD is used to compute the DCC according to (\ref{fou}), (\ref{mf2}), and (\ref{dcc}). Eq.  (\ref{itq44}) is, on the other hand, achieved by interpolating the experimental mass spectrum of \smr (see Fig. \ref{cem44}).

The first element in the neutral strange vector kaon  family to be analyzed is denoted by  $K^*_4$, which corresponds to $n=4$. Substituting it back in Eq. (\ref{itp3}), the DCC evaluation yields 1062.16. Then using it in the left-hand side of Eq. (\ref{itq44}) yields the mass of the neutral  $K^*_4$ to be equal to  $m_{K^*_4}= 1937.67$ MeV/$c^2$. Now, putting $n=5$ in Eq. (\ref{itp3}) and the DCC of the neutral $K^*_5$ \smr reads 1350.01 nat. Afterward, Eq. (\ref{itq44}) can be solved when this value is replaced on the left-hand side of Eq. (\ref{itq44}). The associated mass of  $K^*_5$ reads $m_{K^*_5}= 2107.73$ MeV/$c^2$.  
Correspondingly, the next excited state, regarding $n=6$ can be considered in Eq. (\ref{itp3}), whose evaluation yields the DCC of the $K^*_6$ \smr to equal 1624.96 nat. By solving  Eq. (\ref{itq44}) for it yields the mass  $m_{K^*_6}= 2242.16$ MeV/$c^2$. These outcomes are summarized in Table 
\ref{scalarmasses103}. 
	\begin{table}[H]
\begin{center}\begin{tabular}{||c|c||c|c||}
\hline\hline
$n$ & State & $M_{\scalebox{.67}{\textsc{Experimental}}}$ (MeV/$c^2$)  & $M_{\scalebox{.67}{\textsc{Theory}}}$ (MeV/$c^2$) \\
       \hline\hline
\hline
1 &\;$K^*(892)\;$ & $891.67\pm0.26$ & 1035.2   \\ \hline
2 &\;$K^*(1410)\;$ & $1414\pm15 $ & 1446.9   \\ \hline
3& \;$\;K^*(1680)$& $1718 \pm 18$       & 1749.5     \\\hline
4& \;$K^*_4\;$& ----------  & 1937.67${}^\star$  \\\hline
5& \;$K^*_5\;$& -----------     &  2107.73${}^\star$   \\\hline
6& \;$K^*_6\;$&  -----------   & 2242.16${}^\star$ \\\hline
\hline\hline
\end{tabular}
\caption{Table \ref{scalarmasses1}, with also the  $n=4, 5, 6$ neutral  \smr added. The extrapolated masses  in the fourth column indicated with a `` ${}^\star$ '' indicates the use of both DCC-Regge-like trajectories (\ref{itp3}, \ref{itq44}), 
which interpolate the experimental mass spectrum of \smr from $n=1, 2, 3$. } \label{scalarmasses103}
\end{center}
\end{table}
	The neutral $K^*_5$ strange vector kaon resonance, whose mass is predicted to be 2107.73 MeV/$c^2$, might match the $X(2100)$ further mesonic state in \cite{pdg}, which has experimental mass 2100$\pm$40 MeV/$c^2$ \cite{pdg}. In addition, the neutral $K^*_6$ strange vector kaon resonance, whose mass is predicted to be 2242.16 MeV/$c^2$, might be identified to the $X(2210)$  state in \cite{pdg}, which has experimental mass 2210$_{-21}^{+79}$ MeV/$c^2$ \cite{pdg}.

\subsection{DCC of charged strange vector kaons}
\label{cdcc}
 Now the charged \smr are going to be revisited and probed by the DCC. The first configuration regards the state 
 $K^*(892)^+=(u\bar{s})$, whereas the another possibility is given by 
$K^*(892)^-~=~(\bar{u} s)$, both for $n=1$.
Respectively the same constituent quark models are used to describe the $K^*(1410)^\pm$ and $K^*(1680)^\pm$ strange vector kaons. 
 The average quark mass is provided by $\bar{m} = 411$ MeV/$c^2$, while the deformation parameter in the dilaton is given by ${\scalebox{.94}{$\upalpha$}}= 0.0554$ and  ${\scalebox{.94}{$\kappa$}} = 530.79$ MeV/$c^2$. 

\begin{table}[H]
\begin{center}
\begin{tabular}{||c|c|c||}
\hline\hline
$n$ & State & DCC \\
       \hline\hline
\hline
1\; &\;$K^*(892)\;$ & \;106.98\;   \\ \hline
2\; &\;$K^*(1410)\;$ &\; 179.11\;   \\ \hline
3& \;$\;K^*(1680)$& \;258.15\;          \\\hline
\hline\hline
\end{tabular}
\caption{DCC of charged strange vector kaons.} \label{scalarmasses6}
\end{center}
\end{table}

\begin{figure}[H]
	\centering
	\includegraphics[width=8.9cm]{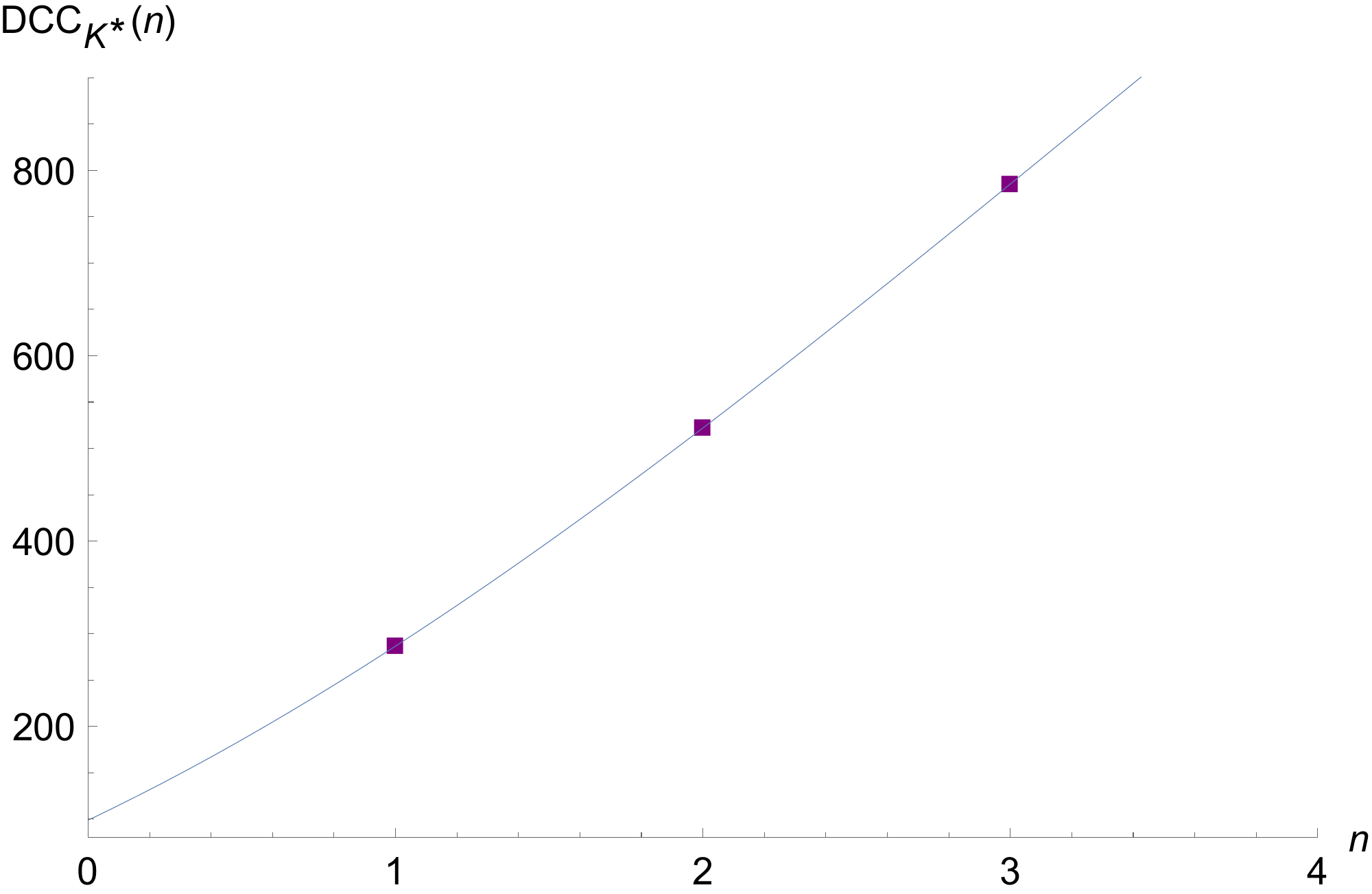}
	\caption{DCC of the charged strange vector kaons as a function of the $n$ radial excitation level, for $n=1,2,3$.  
The first form  of the DCC-Regge-like trajectory is plotted as a dark gray line.}
	\label{cen66}
\end{figure}

The  first form of DCC-Regge-like trajectory regards the DCC as a function of $n$. Fig. \ref{cen66} shows the obtained outcomes, whose cubic polynomial  interpolation of data in Table \ref{scalarmasses6} generates  the first type of DCC-Regge-like  trajectory, 
\begin{eqnarray}\label{itp4}
\!\!\!\!\!\!\!\!\!\!\!\!\clt{{\rm DCC}}_{K^*}\!\!&\!=\!&-3.30162 n^3+33.7147 n^2+157.317
   n+98.9497.  \end{eqnarray}
 \clt{within $0.001\%$ RMSD}.

The DCC of \smr can be  also realized  as a function of mass spectra of the \smae. In this way, the second type of DCC-Regge-like trajectories regards the experimentally detected  mass spectra of the \sma \cite{pdg}. 
\begin{figure}[H]
	\centering
	\includegraphics[width=9cm]{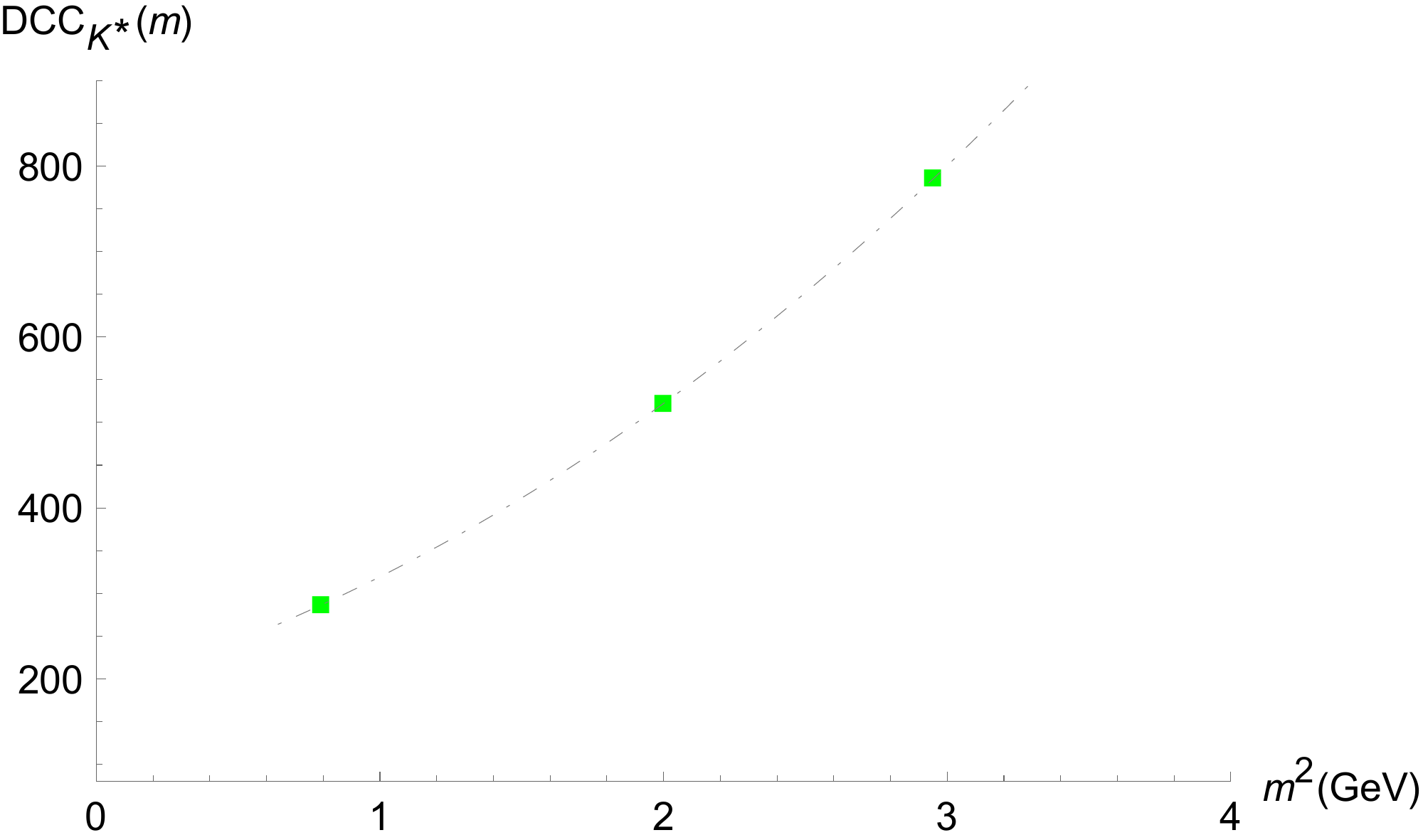}
	\caption{DCC of the \smas family as a function of their squared mass, for  $n=1,2,3$ (respectively corresponding to the $K^*(892)^\pm$, $K^*(1410)^\pm$, and $K^*(1680)^\pm$  resonances in PDG \cite{pdg}). 
The DCC-Regge-like trajectory is represented by the interpolating dashed line.}
	\label{cem44c}
\end{figure}
\noindent The second type of DCC-Regge-like trajectory, relating the DCC of the charged strange vector kaons to their squared mass spectra, $m^2$ (GeV${}^2/c^4$), has the following interpolation formula:
\begin{eqnarray}
\label{itq55}
\!\!\!\!\!\!\!\!\!\!\!\!\!\!\!\!\!\!\clt{{\rm DCC}_{K^*}(m)} \!&\!=\!& -0.382015 m^8+3.54017 m^6+26.0607
   m^4+105.52 m^2+184.683, 
   \end{eqnarray} \clt{within $0.001\%$  RMSD}.    
   
Both the Regge-like trajectories (\ref{itp4}) and (\ref{itq55}) can then be employed to extrapolate the mass spectrum of the \smrs. Eq. (\ref{itp4}) yields the DCC of \sma for any value of the radial excitation $n$. One can therefore switch it on the left side of Eq. (\ref{itq55}) and algebraically solve it for  the variable $m$, yielding the mass spectrum of the \smr for $n>3$. First, we aim to provide the mass of the  charged $K^*_4$ strange vector kaon resonance, with  $n=4$. 
Putting $n=4$ back into Eq. (\ref{itp4}), the DCC amounts to \clt{337.507 nat}. Therefore, using it on the left-hand side of Eq. (\ref{itq55}), one can obtain the  algebraic solution for the $m$ variable, yielding the mass  \clt{$m_{K^*_4}= 1935.34$ MeV/$c^2$}. Now, accomplishing the same technique for $n=5$ in Eq. (\ref{itp4}), the DCC underlying the $K^*_5$ resonance reads \clt{410.590 nat}. Subsequently  working out Eq. (\ref{itq55}) for this value of the DCC implies the  mass  \clt{$m_{K^*_5}= 2099.83$ MeV/$c^2$}.  
Analogously,  the $n=6$ excitation level can be now implemented into Eq. (\ref{itp4}), implying the DCC  to be equal to \clt{470.807 nat}. When Eq. (\ref{itq55}) is then resolved, using this value of DCC, the $K^*_6$ resonance mass  \clt{$m_{K^*_6}= 2223.76$ MeV/$c^2$} is acquired. These results are compiled in Table 
\ref{scalarmasses1004}. 
	\begin{table}[H]
\begin{center}\begin{tabular}{||c|c||c|c||}
\hline\hline
$n$ & State & $M_{\scalebox{.67}{\textsc{Experimental}}}$ (MeV/$c^2$)  & $M_{\scalebox{.67}{\textsc{Theory}}}$ (MeV/$c^2$) \\
       \hline\hline
\hline
1 &\;$K^*(892)\;$ & $891.67\pm0.26$ & 1035.2   \\ \hline
2 &\;$K^*(1410)\;$ & $1414\pm15$ & 1446.9   \\ \hline
3& \;$\;K^*(1680)$& $1718 \pm 18$       & 1749.5     \\\hline
4& \;$K^*_4\;$& ---------  & 1935.34${}^\star$  \\\hline
5& \;$K^*_5\;$& -----------     & 2099.83${}^\star$   \\\hline
6& \;$K^*_6\;$&  -----------   & 2223.76${}^\star$ \\\hline
\hline\hline
\end{tabular}
\caption{Experimental and AdS/QCD mass spectrum (for $n=1,2,3$), with the higher $n$  resonances (with an asterisk) of the strange vector kaon meson family. The extrapolated masses for $n=4, 5, 6$ in the fourth column indicated with a `` ${}^\star$ '' specify that they are the result of the concomitant use of DCC-Regge-like trajectories (\ref{itp4}, \ref{itq55}), interpolating the experimental masses from $n=1, 2, 3$ resonances. } \label{scalarmasses1004}
\end{center}
\end{table}
 Comparing the predicted mass spectrum values using the DCC approach for charged and neutral strange vector kaons, a relatively small percentage difference is obtained: the value is 0.13\% for the $K^*_4$ resonance, and  the difference increases to 0.37\% in the case of the $K^*_5$ resonance, whereas this value is 0.82\% relative to the $K^*_6$ resonance. Analogously to the mass spectra of charged and neutral \smr in the DCE approach, these percentage variations are very slight. It reveals that 
 the mass spectrum of the next generation of charged and neutral \smr 
are quite similar, providing  a robust perspective for future  analyses involving experimental data. Besides, the $K^*_6$ strange vector kaon resonance, whose mass is  predicted to be 2203.43  MeV/$c^2$, might match the  $X(2210)$ further mesonic state in \cite{pdg}, which has experimental mass  $2210^{+79}_{-15}$ MeV/$c^2$ \cite{pdg}.

\section{Concluding remarks and perspectives}\label{iv}

 DCE and DCC Regge trajectories represent a successful approach to mass spectroscopy, using   AdS/QCD. Here the mass spectroscopy of neutral and charged \smr was thoroughly investigated in this scenario, bringing relevant physical features of the next generation of strange vector kaon radial resonances. Their underlying information entropy content played an important role in deriving the mass spectra of the next generation of strange vector kaon resonances, with  higher radial quantum numbers. To accomplish it, a shift (\ref{devia}) of the quadratic dilaton  was used. One of the advantages of using  DCE- and DCC-based techniques is that  the computational apparatus is straightforward when compared to other phenomenological approaches of AdS/QCD, including the lattice one. Another purpose of the DCE and DCC approaches to \sma is the interpolation of their experimental mass spectrum of already experimentally established \sma in PDG \cite{pdg}, to derive the mass spectra of the next generation of \smrs. 
 The techniques involving the DCE and the DCC in the soft wall AdS/QCD with deformed dilaton took into account the experimental mass spectrum of the strange vector mesons $K^*(892), K^*(1410),$ and $K^*(1680)$. Both charged and neutral strange vector meson families were scrutinized, yielding quite similar results, leading to robust data. This was implemented by the concomitant analysis  of the first and second kinds of DCE- and DCC-Regge-like  trajectories, extrapolating them for $n=4,5,6$, from the interpolation curves.  
 %Our analysis was split into two cases, involving two quark constituent models: the charged  first strange vector kaon resonances,  $K^*(892)^+=(u\bar{s})$ and the $K^*(892)^-~=~(\bar{u} s)$, and the neutral first strange vector kaon resonance  $K^*(892)^0=~(d\bar{s})$. The slightly different mass brings tiny corrections \bltx{Discuss here.} 
The DCE- and DCC-based procedure 
can address the question of predicting  the mass spectrum of the next generation of strange vector kaon resonances. In addition, these methods are more succinct when compared to the predictions of the soft wall AdS/QCD, involving the eigenvalues of the Schr\"odinger-like equations.
It is a complementary option to solving Eq. (\ref{schr}) in order to obtain the mass spectrum of strange vector kaon resonances. Besides, it is a more phenomenological technique, since the DCE-Regge-like trajectories are constructed upon the experimental mass spectrum of the $K^*(892), K^*(1410),$ and $K^*(1680)$ \smre. In this way, experimental data is also being used, and employing the stress-energy-momentum  tensor (\ref{t002}) of AdS/QCD, the DCE was evaluated following the protocol  (\ref{fou}) -- (\ref{confige}). 
Extrapolation of the two kinds of DCE-Regge-like trajectories yielded the mass spectrum of the next generation of \smrs. Two analyses were implemented in Subsecs. \ref{ndce} and \ref{cdce}, using the DCE to and the masses of the charged and the neutral strange vector kaon $K^*(892)$ to predict the mass spectrum of more unstable resonances. The $K^*_4$ mass prediction differed by 0.03\%, whereas the $K^*_5$ resonance has a difference of the predicted mass spectrum consisting of 1.2\%. Comparing once more the charged and the neutral $K^*(892)$ models in Subsecs. \ref{ndce} and \ref{cdce} yields a difference of 1.59\% in the mass of the $K^*_6$ \smrs. These results come from the comparison of Tables 
\ref{scalarmasses102} and \ref{scalarmasses100}. 
A similar analysis encompassing the charged and neutral $K^*(892)$ appears in Subsecs. \ref{ndcc} and \ref{cdcc}, but from the perspective of the results obtained by the DCC approach, which resulted in small differences in the mass spectrum. For the mass of the $K^*_4$ resonance, the difference was 0.13\%, while the amount differed by 0.37\% for the $K^*_5$ resonance and 0.82\% for $K^*_6$.

Looking at the set of results obtained for DCE (Sec. \ref{sec2}) and DCC (Sec. \ref{sec21}), the two procedures can be compared
for charge kaons. The first case involves neutral strange vector kaons, where a difference of 0.20\% exists,  between the results of DCE and DCC for the $K^*_4$ resonance, followed by a difference of 0.72\% for the $K^*_5$ resonance. For the $K^*_6$ resonance, the difference gaps to 1.76\% between the results of the two procedures. In the case of charged strange vector kaons, there is another growth pattern in the percentage difference between the results obtained through DCE and DCC, as a function of $n$. The $K^*_4$ resonance appears with a difference of 0.11\% between the two methods, while $K^*_5$ differed by 0.86\% and $K^*_6$ by 0.68\%. In the DCE approach of neutral strange vector mesons here studied, the $K^*_5$  resonance may either match the  $X(2075)$ or the $X(2080)$ or even the $X(2100)$ mesonic state in PDG \cite{pdg}. 
For the charged strange vector mesons, the $K^*_5$ resonance may match only the $X(2100)$, as well as in the DCC approach for neutral strange vector meson resonances. In all the DCE and DCC approaches for neutral and charged strange vector kaons, the $K^*_6$ resonance might match the  $X(2210)$ mesonic state \cite{pdg}.

In the fifth column of each one of Tables  \ref{scalarmasses1} and  \ref{scalarmasses10}, one can see the relative errors between the mass spectrum of AdS/QCD, obtained as eigenvalues of the Schr\"odinger equation, and the experimental mass spectrum \cite{pdg,ALICE:2021xyh}. 
The DCE-based techniques here studied may reduce these errors, for $n>3$, improving the mass determination of strange vector kaon radial resonances, based on extrapolating the DCE-Regge-like behavior.  Our proposed complementary analysis is that the DCE method here prescribed can provide other relevant phenomenological aspects not provided by the  bottom-up AdS/QCD. 
%The comparison between 
% AdS/QCD-obtained mass spectra and the DCE-based procedure that here hybridizes AdS/QCD should be also analyzed with  experimental data, of the high excitations that are still beyond the current detection threshold, in hands. We hope that soon the running experiments will give us the final answer. 

The DCE-based apparatus that  hybridizes AdS/QCD can also address  the measurement of the degree of chaos in any system. In fact, 
Ref. \cite{Ma:2018wtw} demonstrated that information entropy corresponds to information chaoticity carried by field in QCD \cite{Ma:2018wtw}.  The chaotic profile of high-energy collisions that produce \smr  can be then estimated by the depletion of information entropy in the final stages of  particle collisions. These quantitative aspects are therefore read off by the DCE. 
Strange vector meson resonances are experimentally produced in multiparticle production processes, as in the observation of charged strange kaons \cite{ALICE:2014jbq} and the recent results on $K^*(892)^\pm$ resonance production \cite{ALICE:2021xyh}, both of them obtained by the ALICE detector in heavy-ion and proton-proton collisions at high energies. 	Among the hadrons produced in these processes, the resonances $K^*$ and $\bar{K}^*$ play a sensitive role as probes capable of bringing information from stages before the hadronization phase of mediums with high levels of temperature and density, such as the quark-gluon plasma, where we find unconfined quarks \cite{Ilner:2016xqr}. Analyses of momentum distribution in reactions of involving $\pi^-$ scattering  show a high sensitivity of the meson $K^*(892)^+$ to changes in the in-medium mass shift in the low-momentum region, which highlights a way to discover properties of the medium \cite{Paryev:2020ivs}.

 Another important feature of this method is the configurational stability of the \smrs. In Tables \ref{scalarmasses50} and \ref{scalarmasses60}, respectively for neutral and charged strange vector kaons, the values of the DCE of higher excitation level resonances are shown  to increase monotonically, according to $n$. This implies a configurational instability for higher excitation level \smrs. It also corroborates the fact that lower excitation level resonances have been already detected in experiments \cite{pdg}, as they are more dominant when compared to states with the higher quantum number $n$. The less massive the  strange vector kaon state, the less configurationally unstable they are. Under the point of view of the DCC, \smr with  lower $n$ have less complexity. The extrapolation of the mass spectrum of charged and neutral \smr can be, at least numerically,  implemented for any excitation number $n$, with good accuracy. However, our analysis considered three resonances with quantum numbers $n=4,5,6$. Beyond that, further vector kaon states are not explored, since they present too high values of DCE and DCC, being exceedingly unstable and implausible to be detected in current experiments.
  The large values of the DCE  evince a phenomenological lower predominance of \smr with $n>6$.

\medbreak
\subsubsection*{Acknowledgments}  RdR~is grateful to The S\~ao Paulo Research Foundation FAPESP (Grant No. 2021/01089-1 and No. 2022/01734-7) and the National Council for Scientific and Technological Development -- CNPq (Grant No. 303390/2019-0), for partial financial support. PHOS thanks to Coordination for the Improvement of Higher Education Personnel (CAPES - Brazil).

\bibliography{bibliography}

\end{document}